\documentclass[a4paper,11pt]{article}
\pdfoutput=1

\usepackage{jinstpub}
\usepackage{graphicx}
\usepackage{amssymb}

\title{\boldmath The FoCal prototype -- an extremely fine-grained electromagnetic calorimeter using CMOS pixel sensors}

\author[a]{A.P.~de Haas,}
\author[a,1]{G.~Nooren,\note{Corresponding author.}}
\author[a]{T.~ Peitzmann,}
\author[a]{M.~ Reicher,}
\author[a]{E.~Rocco,}
\author[b]{D.~R\"{o}hrich,}
\author[b]{K.~Ullaland,}
\author[a]{A.~van den Brink,}
\author[a]{M.~van Leeuwen,}
\author[a]{H.~Wang,}
\author[b]{S. Yang,}
\author[a]{C. Zhang}

\affiliation[a]{ Institute for Subatomic Physics Utrecht University and Nikhef, P.O.B. 80000, 3508 TA Utrecht, the Netherlands}
\affiliation[b]{Institute for Physics and Technology, University of Bergen, Allegaten 55, 5007 Bergen, Norway}

\emailAdd{nooren@nikhef.nl}
\arxivnumber{}

\abstract{A prototype of a Si-W EM calorimeter was built with Monolithic Active Pixel Sensors as the active elements. With a pixel size of 30 $\mu$m it allows digital calorimetry, i.e. the particle's energy is determined by counting pixels, not by measuring the energy deposited. Although of modest size, with a width of only four Moliere radii, it has 39 million pixels.
In this article the construction and tuning of the prototype is described. Results from beam tests are compared with predictions of GEANT-based Monte Carlo simulations. The shape of showers caused by electrons is shown  in unprecedented detail. Results for energy and position resolution are also given.
}

\begin{document}
\maketitle

\section{Introduction} \label{sec:intro}

Traditionally, electromagnetic calorimeters have a granularity which is of the order of magnitude of the Moli{\`e}re radius of the detector material. This coarse sampling of the transverse shower shape is sufficient to discriminate electromagnetic and hadronic showers. The transverse position of the shower maximum as an estimator of the position of impact of the incoming particle can be calculated from the centre of gravity of the signals' distribution in the detector cells. The resolution obtained by such methods is typically an order of magnitude smaller than the cell size (a few mm for cells with a transverse size of a few cm).

While such a granularity is usually sufficient for the measurement of well separated individual particles (photons and electron/positrons) in state-of-the-art high-energy physics experiments, high particle densities as encountered at the highest beam energy e.g. in high-energy jets and in particular at high rapidities, i.e. small polar angles with respect to the particle beams at accelerators, pose new challenges. So far, jet measurements have mostly been performed by summing calorimeter energies from cells within phase space areas, so-called \textit{jet cones}. It has been demonstrated, however, that the knowledge of the particle composition in the jet is crucial for reducing the jet energy scale uncertainties. \textit{Particle flow algorithms} are expected to make use of the capability to track individual particles in a calorimeter \cite{pfa}. Such algorithms benefit enormously from a higher granularity of the calorimeter.

 \textit{Direct} photons emitted in high-energy nuclear collisions can probe the initial state (e.g. parton distribution functions), and thermal photons are among the most interesting signals of the quark-gluon plasma \cite{peitzmann:2002}. However, the measurement of high-energy photons is a challinging task: decay photons, in particular from $\pi^0$ decays, constitute a large background in an inclusive photon measurement. This background can be reduced by e.g. isolation cuts and also by a direct rejection of photon pairs matching the decay kinematics. But the latter discrimination method has to fail for high energy $\pi^0$ mesons, when the two showers from the individual decay photons merge. Some discrimination is still possible via an analysis of the shower shape, but the efficiency of this method will also decrease substantially at very high energy.

The granularitiy of the detector is the main limiting factor for two-shower separation and shower shape analysis. The transverse shower shape is very strongly peaked around the shower centre, so that granularities much smaller than the Moli\`ere radius should prove to be valuable. Such higher granularity will in addition greatly improve the position resolution of calorimetric measurements. Some of the more recent calorimeter implementations do in fact partially use higher granularity layers (strip detectors) to exploit these possibilities, as in the CMS pre-shower detector \cite{cms}, and full three-dimensional shower reconstruction was used in the PAMELA space mission \cite{pamela}. 

The measurement of direct photons at forward rapidities is expected to be sensitive to the effects of gluon saturation.
The ALICE experiment is considering such a measurement, while aware that this will be extremely difficult and would require an update of the existing setup \cite{peitzmann:2014}. The studies presented in this paper are performed in the context of R\&D for a proposed upgrade of ALICE with a forward calorimeter (FoCal) based on a Si-W sandwich structure. Monolithic Active Pixel Sensors (MAPS) \cite{Turchetta} is considered to be the most promising sensor technology to provide the required high granularity. In such a detector calorimetric energy would be measured via the number of pixels above a predefined signal threshold. For a \textit{digital calorimeter} to work the average occupancy per pixel should be $\ll 1$. With a particle density of 10$^3$ mm$^{-2}$ or larger in the shower core, one requires very small pixels of 50-100~$\mu$m at the most. Unfortunately, while in principle the processes contributing to electromagnetic showers are known theoretically, there is little experimental knowledge on the details of shower development on such small scales, and the design of detectors so far relies mainly on Monte Carlo simulations.

For the related R\&D studies, a prototype of a Si-W calorimeter with extremely high granularity and very small Moli\`ere radius has been designed and constructed. The main motivations for building this prototype were
\begin{itemize}
\item to demonstrate the feasibility of a pixel counting digital calorimeter,
\item to explore the possibilities offered by MAPS sensors in such an application, and
\item to obtain detailed information on electromagnetic shower development on the scale of $\approx100~\mu$m.
\end{itemize}
After introducing the design of this prototype, its implementation in a Monte Carlo simulation program will be described followed by the correction and analysis procedures performed on the data. Finally, results will be shown from beam test measurements and compared to simulations. Earlier descriptions of the prototype and some preliminary results can be found in \cite{RD11,peitzmann:2013,calor,martijn}.

\section{Prototype Design} \label{sec:des}

A calorimeter based on MAPS was proposed in the context of the CALICE project for ILC (see \cite{Dauncey3} and references therein) and a crude prototype was even tested with beam \cite{price},  but this is the first implementation of a full MAPS calorimeter prototype. 
In the last decade a vigorous development of MAPS has occurred at a number of different locations, like RAL\cite{sensors}, 
IPHC (MIMOSA \cite{voutsinas}), CERN \cite{cern} and SLAC \cite{segal}, mainly in view of upgrades of the trackers of STAR \cite{star} and ALICE \cite{ITS}. 
Still, for the present purpose the choice of available sensors was limited, as a reasonably large size is required. In addition one needs to have the possibility to read out all pixels, because the occupancy is expected to vary strongly; -- and data reduction as implemented in many such chips aimed at tracking applications is not suitable.

\subsection{Mechanics}

The PHASE2/MIMOSA23 from IPHC \cite{phase1} was chosen.
This is the only full reticle size MAPS ($640 \times  640$ pixel matrix), which allows the continuous readout of all pixels,
thanks to four outputs at 160~MHz. The 1~MHz rolling shutter corresponds to an integration time of 640~$\mu$s.
The resulting low event rate without pile-up makes this unusable for modern particle physics experiments, but this poses no problem in the case of test measurements. The small pixel pitch of 30~$\mu$m allows very fine sampling of the shower core.

An engineering run at AMS\footnote{AMS-C35B4 OPTO by Austria Microsystems.} provided 5 wafers with a
high resistivity (400~$\Omega$cm) epitaxial layer of 15 and 20~$\mu$m thickness
and 1 standard wafer (10~$\Omega$cm, 14~$\mu$m). The former were thinned down to 120~$\mu$m, the latter to 180~$\mu$m. Because of the relatively low yield, the detector had to be built with different types of sensors, also including the standard wafer.

Compared to a conventional sensor, the active pixel sensor dissipates much more heat, typically 
0.1~W$\cdot$cm$^{-2}$. It was realised that the heat can be transported from the sensor chips to the outside
 by using the rather good heat conductivity of pure W, namely 170~W/m/K (Al: 220~W/m/K)
\footnote{The use of the W absorber as cooling element was also mentioned in proposals for an ILC analogue \cite{frey} and digital \cite{Dauncey4} calorimeter.}.
 Thus the W-absorbers can also serve as heat conductors connected to cooling elements at their edges. The absence of a separate layer of cooling elements leads to a very compact calorimeter, helping to achieve a small Moli\`ere radius.

The detector consists of 24 layers made of pure tungsten absorbers, silicon sensors, printed circuit boards (PCB) and glue . The total thickness of a single layer is 4~mm, 3~mm of which consist of tungsten. The resulting radiation thickness of one layer is 0.97~$X_0$. The use of MAPS allows the sensor part of the layer, including the PCB responsible for readout, to be kept as thin as 1~mm, which leads to a very small Moli\`ere radius, calculated to be $R_\mathrm{M} \approx 11 \, \mathrm{mm}$. 

\begin{figure} [tbh]
\centering
\includegraphics[width=\textwidth]{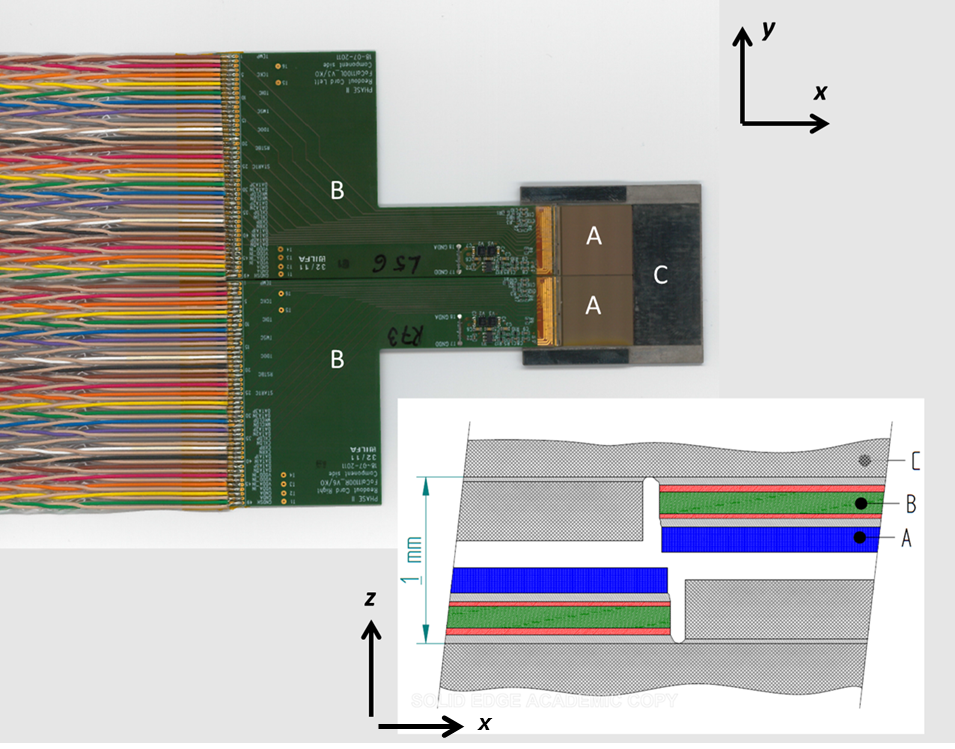}
\caption{Top view of a module showing two sensors \textit{A}  bonded onto two PCBs \textit{B}  side-by-side on the tungsten absorber \textit{C}. Two such modules are combined to form a single layer of the prototype. \textit{Insert}: Schematic cross-section of the $x-z$ plane of the centre part of a layer showing the overlap and the small air gap between the top and the bottom modules. The thickness of each absorber is 1.5~mm. The filler plate (grey in the figure) is 0.3 mm. The total thickness of a layer is 4~mm.
}
\label{fig:setup}
\end{figure}

The active area of a layer is $4 \times 4$ cm$^2$, composed of four sensors, 
while the absorber measures $5 \times 5$ cm$^2$.
Besides the pixel matrix, the MIMOSA23 contains discriminators and control and output circuitry.
 Because of this, the total sensor chip  contains some insensitive areas, so it was decided to let these overlap in one transverse direction to minimise dead areas in the detector. 
For ease of construction there still remains a dead zone of  0.1 mm between each pair of chips in the other direction.
A layer is composed of two identical modules, each with two sensors, one mounted upside down on the other, see figure~\ref{fig:setup}.
In this way the cables stick out in opposite directions, allowing a compact design.
A filler plate of 0.3~mm of tungsten is glued next to the sensors in order to reduce the amount of low-$Z$ material\footnote{Although in this way the sensors at negative $x$ have 7\% more radiation depth in front than the sensors at positive $x$, no effect was seen for the full detector. This is because the effect is not cumulative: the total radiation depth is equal.}.
The free sides of each absorber are pressed against copper cooling elements.
This design results in a gap in the $y$-direction of 0.4 mm and an overlap in the $x$-direction of 0.1 mm.
These are nominal distances between the pixels in adjacent sensors. Actual distances are determined by the alignment procedure.

Taking into account the Moli\`ere radius, the tower is wide enough to fully contain showers and to study the lateral shower development.
The first active layer (\textit{layer 0}) has an aluminium absorber without filler plate (total 0.02~$X_0$) in front, to act as a charged particle detector.
Between layers 21 and 22 6.7~$X_0$ of tungsten alloy (Densimet 18) are placed to obtain a total depth of 28~$X_0$.
 
\begin{figure} [tbh]
\centering
\includegraphics[width=0.8\textwidth]{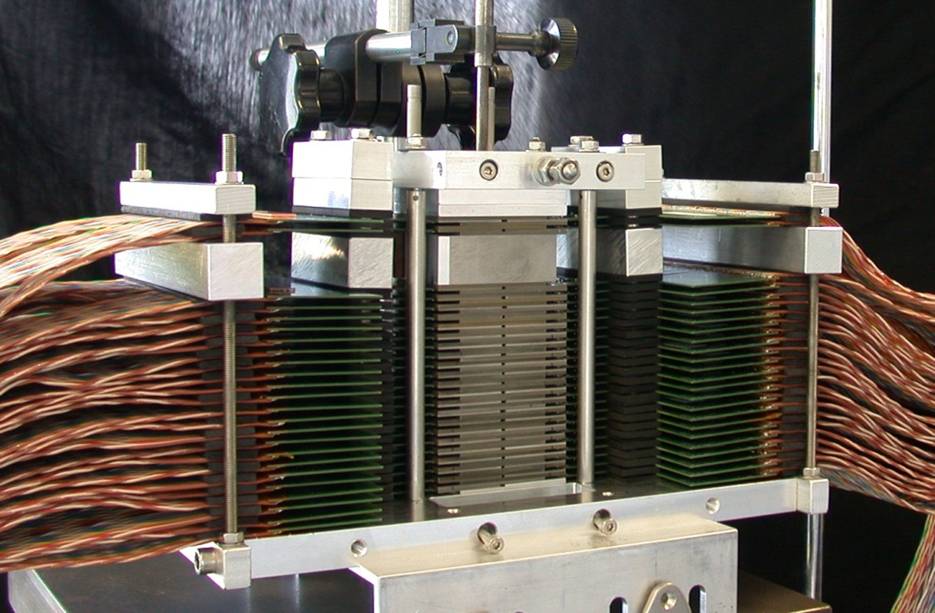}
\caption{Side view of the prototype detector, without the cooling system. The beam direction is from below ($z$ axis points upward). The total length in the beam direction is 11 cm, made up of 22 layers,
a 20 mm W absorber and 2 additional layers. On both sides of the greyish stack, green PCBs with flat cables can be seen, reading out 1 sensor each. 
}
\label{fig:prototype3D}
\end{figure}

Figure~\ref{fig:prototype3D} shows the detector with its main components, but without the cooling system.
The printed circuit boards (PCBs) visible in this figure extend into the tower and connect the sensor chips to the flat cables.
The total detector counts 96 sensors in 24 layers.
The coordinate system is indicated in figure~\ref{fig:setup}. Each sensor is defined by the quadrant $q$ and the layer $l$.
The first layer ($z =0$) has $l=0$. Quadrants are numbered clockwise with $q=0$ for $x>0$ and $y>0$.

\subsection{Electronics}
\subsubsection{Tuning}  \label{sec:tuning}

Due to diffusion, the charge created by a particle will lead to a cluster of pixel hits.
The size of the cluster will depend on the charge created by the particle, the charge diffusion and collection, and the threshold of the discriminators.
For application in trackers the discriminators are usually adjusted such that an acceptable fake rate,
measured as clusters not belonging to a track, is achieved.
In the case of a calorimeter it is a priori not known whether the clusters will be well-separated,
especially in the core of the shower.
In fact, as can be seen in the example of figure~\ref{fig:coredensity}, extremely large clusters can appear in these regions. 
For these cases a simple cluster number will not be an appropriate signal of the deposited energy.
Therefore it was decided to use the number of \textit{hit pixels} as a measure of the energy,
instead of trying to derive the number of \textit{particles} from the hit distributions.\footnote{A similar approach was used by CALICE DHCAL, albeit with much larger (1~cm) pixels.\protect\cite{DHCAL}}

Each sensor has 640 discriminators, one for each column of 640 pixels. 
All the discriminator thresholds are set via two 8-bit DAC values, where one regulates the uniformity within the sensor (i. e. along the $y$-axis) and the other is responsible for the absolute threshold value \cite{phase1}.
Any variation in sensitivity in the $x$-direction cannot be compensated.
The tuning procedure consists of scanning this two dimensional parameter space until the desired noise level is obtained with the lowest spatial variations. 

\begin{figure} [tbh]
\centering
\includegraphics[width=0.4\textwidth]{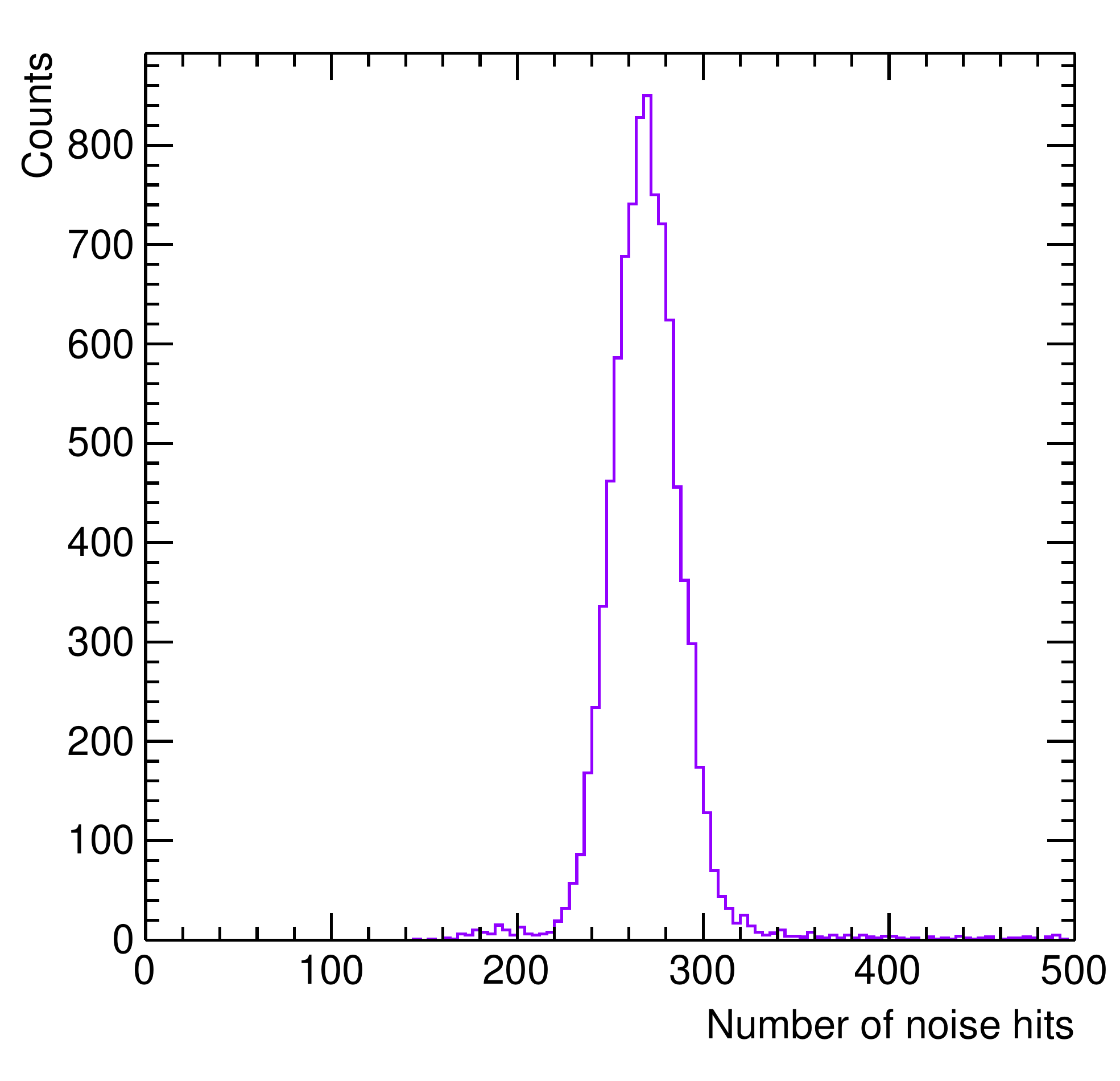}
\caption{Histogram of the number of noise hits in the calorimeter. The first layer, which acts as a charged particle detector, is not included. 
}
\label{fig:noise}
\end{figure}

Some ($ < 1$ \%) pixels show excessive noise which cannot be suppressed by higher discriminator thresholds. These \textit{hot pixels} were disabled via a software mask applied to the raw data. As a result a noise hit fraction of 10$^{-5}$/pixel is obtained.
For the first layer the threshold was set lower to increase the single particle detection efficiency. The resulting noise hit fraction was 10$^{-4}$/pixel.
The noise spectrum of the calorimeter, without the charge particle detection layer 0, is shown in figure~\ref{fig:noise}.
The peak can be fitted with a Gaussian at position 269 and a $\sigma = 17$.
Pedestal (noise) data were collected in between beam spills (SPS) or with the beam stopper in (DESY).

Note that the tuning procedure does not guarantee equal sensitivity of the sensors and calibration is required, see section~\ref{sec:shower}.
 Some sensors are not functioning correctly (e.g. communication errors) and are (partially) turned off. 
These dead sensor areas, together with the pixel area masked because of high noise, amount to an insensitive fraction of $\approx 17 \%$ of the total sensor area installed.

\subsubsection{Readout and Data Processing} \label{sec:read}

\begin{figure}  [tbp]
\centering
\includegraphics[width=\textwidth]{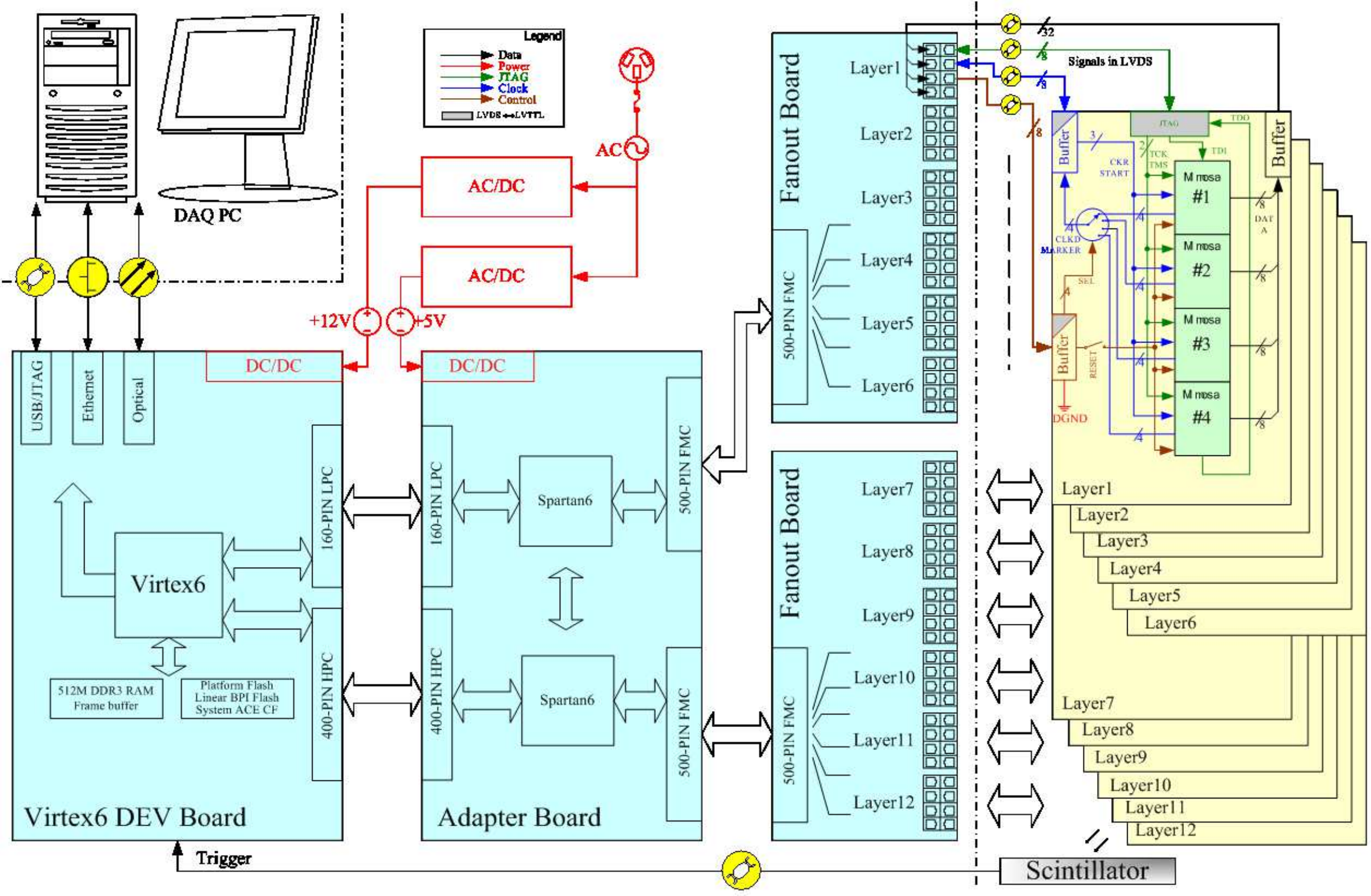}
\caption{Schematic of the read-out electronics for the prototype tower. Only the circuits for 12 layers are shown.}
\label{fig:readoutshiming}
\end{figure}

A data frame of 640 lines with 640 pixels per sensor contains 39 M pixels, so the raw data rate of the prototype is 61~Gb/s.
Several FPGAs are used to manage this, as described in \cite{fehlker}.
A schematic for one half of the prototype is shown in figure \ref{fig:readoutshiming}.
Up to four trigger signals are time-stamped with the read-out clock and stored separately.
A 4~GByte local buffer memory allows to store 0.55~s of pixel and trigger data.
Transfer to the DAQ computer takes $\approx$100 s due to the implementation of the TCP/IP protocol.
Upon completion, the DAQ system waits for the next spill signal to store another 0.55~s of data.
Note that the rolling shutter and chip read-out clocks are running continuously 
to keep a stable phase relation between all chips.
After a full read-out cycle of 640 lines two lines with synchronisation patterns containing a unique ID are added by the sensor chip, making the total cycle 642~$\mu$s.

Each sensor chip performs two steps of multiplexing -- the 640 discriminators connect to 16 intermediate channels, and finally to 4 output channels read out at 160~MHz. The signals of 24 sensors (i.e. 96 channels) are combined by one Spartan6 XC6SLX150 FPGA, and the output of two such Spartan FPGAs is read by one Virtex6 XC6VLX240T FPGA. Two Virtex FPGAs running in a master-slave relationship are needed to read out all sensors and send the data to the DAQ computer. As a pre-processing step before analysis, the collected RAW data then have to be de-multiplexed to recover the true frame data structure. In this step checks on data integrity and synchronisation are also performed. 
From this continuous data stream the off-line processing reconstructs frames consisting of pixels 
read out up to 642~$\mu$s after each trigger 
-- this provides all relevant detector signals corresponding to one event.
Finally, the synchronisation lines are removed from the frames.

\subsection{Measurement Setup}

\subsubsection{Muons from Cosmic Rays} \label{sec:cosmic}

The detector stack was mounted on a metal rail together with two scintillators to be used for triggering. The front (F) and back (B) scintillators have a transverse size of $40 \times 40 ~\mathrm{mm}^2$ and are mounted such that they cover the sensitive region of the calorimeter. The 96 PCBs of all sensors are connected to the DAQ system via flat cables attached to the sides (partially visible in figure~\ref{fig:prototype3D}). The scintillator PMTs are connected to discriminators to provide trigger signals. The logical signals of the trigger are also sent to the DAQ. This setup was mounted vertically for the measurement of muons originating from cosmic rays.

\subsubsection{Beam Tests}

Test beam measurements performed at DESY and at the CERN SPS\footnote{Earlier tests were performed in 2012 with a slightly different detector. The results are published in the Ph.D. Thesis of M.~Reicher \cite{martijn}.} have been used in this analysis.
The corresponding data samples are summarised in Table 1.

\begin{table}[h]
\caption{\protect\label{tab:datasamples} The properties of the different data samples collected in test beams.
}
\begin{center}
\begin{tabular}{|c|l|l|l|l|}
\hline
 site & year & particle type & $p$~(GeV/$c$) & $N_\mathrm{events}$ \\ \hline
 DESY T22 & 2014 & $e^+$ & 2-5.4 & 9.5 $\cdot 10^3$\\ \hline
 CERN SPS H8 & 2014 & $e^-$(mixed) & 244  & 1.6 $\cdot 10^4$\\ \hline
 CERN SPS H8 & 2014 & $e^+$(mixed) & 30, 50, 100 & 3 $\cdot 10^4$,  3 $\cdot 10^4$, 8 $\cdot 10^4$\\ \hline
\end{tabular}
\end{center}
\label{default}
\end{table}

\begin{figure}  [tbp]
\centering
\includegraphics[width=0.45\textwidth]{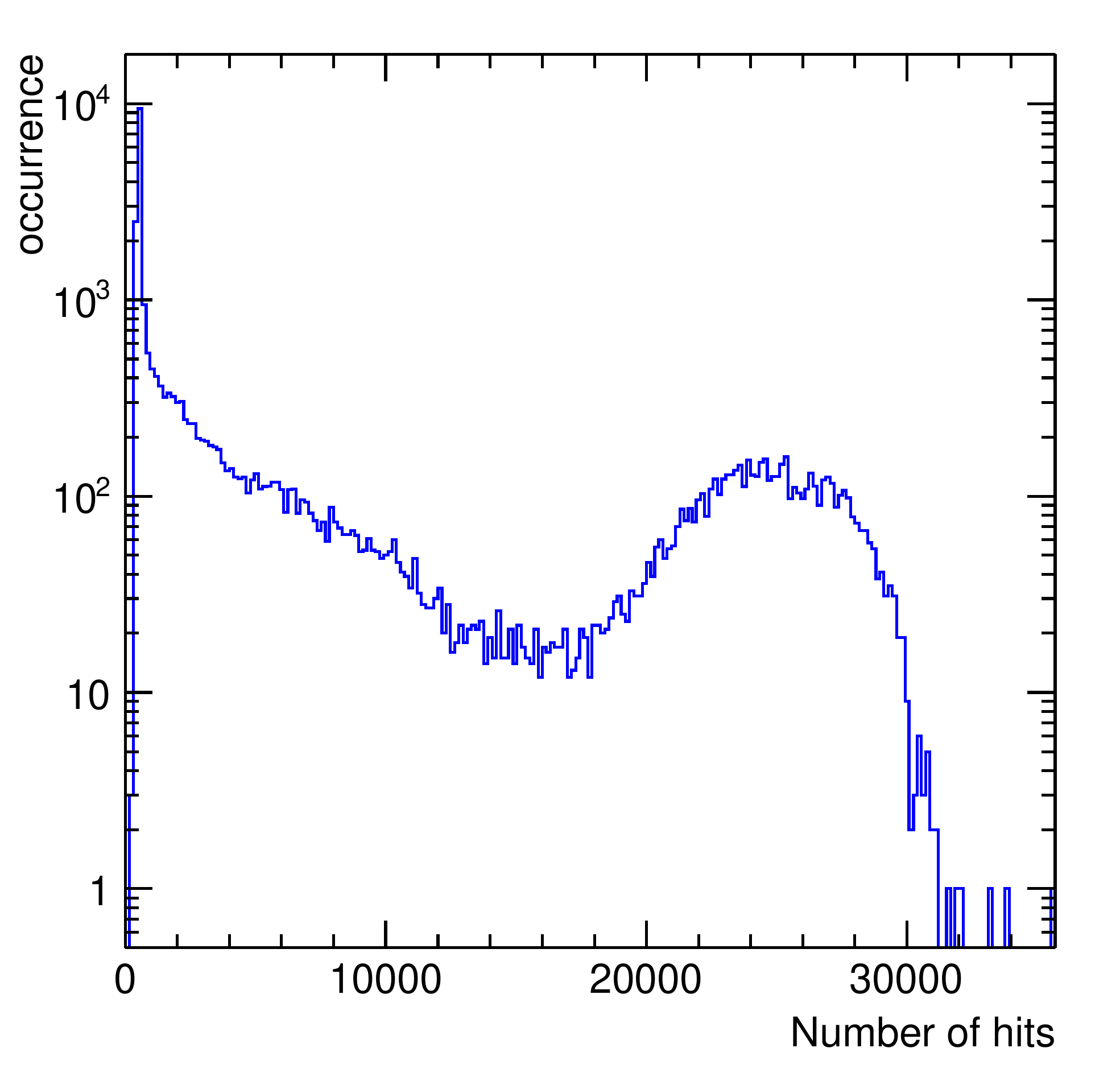}
\includegraphics[width=0.47\textwidth]{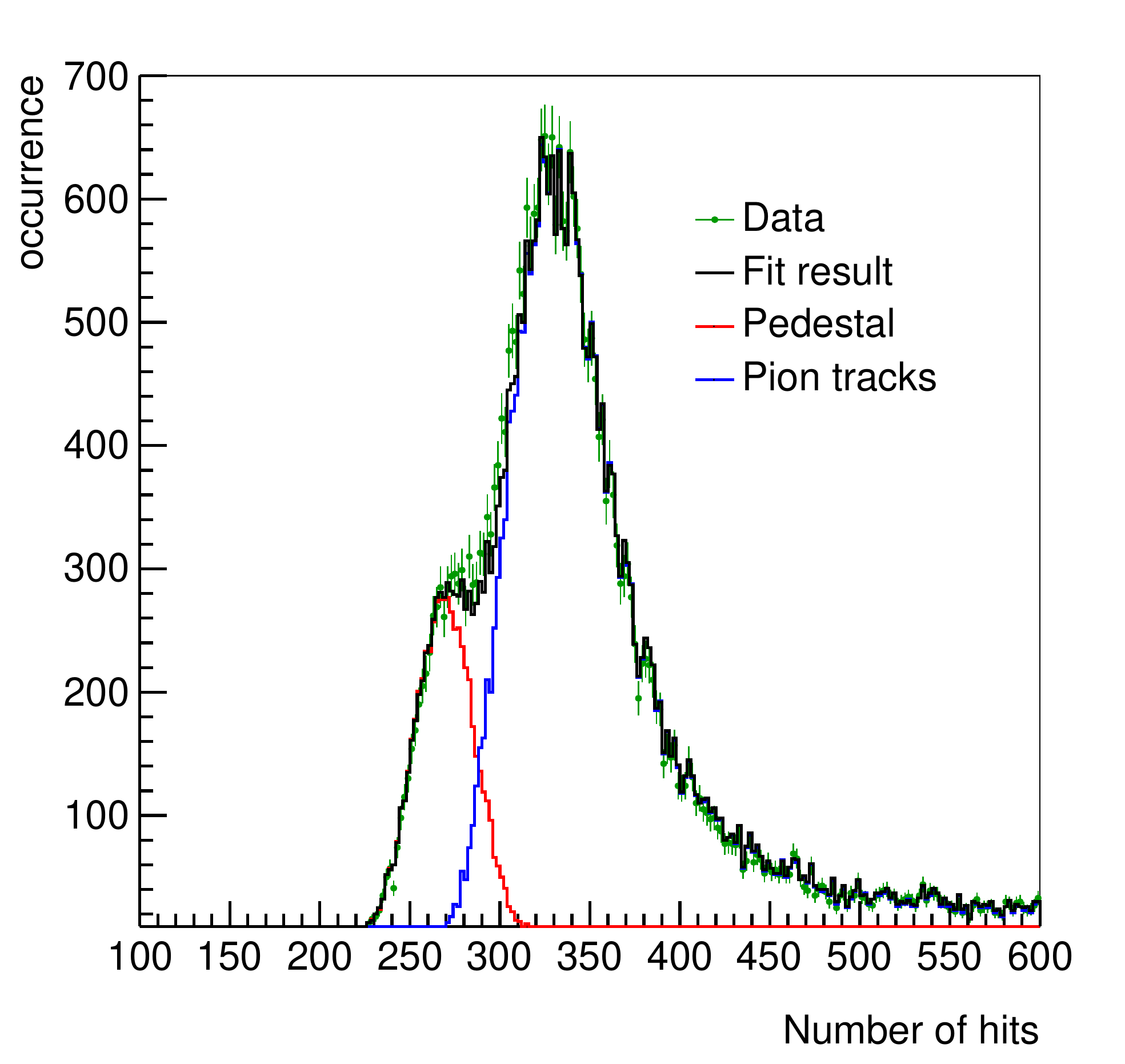}
\caption{Histogram of the number of hit pixels at 100~GeV/$c$. Uncalibrated data (82 k events). Left: Full range, the MIP peak and full energy peak are easily visible. The entries in between are due to partially developed (hadronic) showers.  Right: the region around the MIP peak with a fit of appropriately scaled templates of noise (pedestal) and MIPs (from tracks).
}
\label{fig:histo100GeV}
\end{figure}

 The setup is similar to the one for cosmic measurements, however the orientation of the detector is horizontal. In addition to the F and B scintillators, three more scintillators are available for triggering purposes: a large scintillator (P) in front of the whole setup and two scintillator fingers of 1~cm width oriented horizontally (H) and vertically (V). All trigger signals are defined as coincidences of the F scintillator with at least one other scintillator. 
The data used in this analysis were triggered by H \textsc{and} V \textsc{and} F. In the SPS tests no external detectors were available to help with the particle identification. 
Figure~\ref{fig:histo100GeV} shows the histogram of all hits taken at 100 GeV/$c$.
The broad peak in the left figure around 24000 corresponds to the full energy signal. 
The peak at a few hundred hits contains the pedestal and the signal of minimum-ionising particles (MIPs).
Figure~\ref{fig:histo100GeV}~right shows an enlargement of this peak with fits of the experimental templates of the pedestal (as in figure~\ref{fig:noise}) and MIPs (from reconstructed pion tracks).
From these data one can conclude that the beam contains 76\%  pions.

\section {Simulation} \label{sec:simul}

The detector has been implemented in detail in a setup for GEANT4\footnote{geant4 version 10-00-patch-01 with emStandard} simulations. 
This setup provides energy depositions of particles in the sensitive layer of the silicon sensors for all individual pixels. 
The charge diffusion is modelled by redistributing the total charge equivalent to the energy deposition over several pixels using the model from \cite{difmodel}.
As the electric field is very small, charge is assumed to diffuse isotropically while recombination takes place. 
Charge moving towards the substrate will be reflected by the potential barrier.

\begin{figure}  [tbh]
\centering
\includegraphics[width=0.5\textwidth]{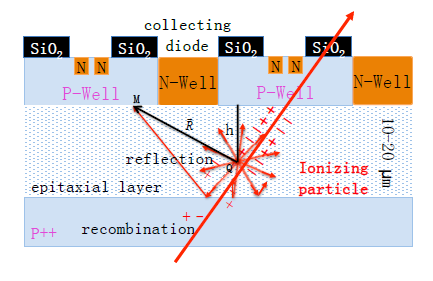}
\caption{The charge diffusion geometry in one dimension.}
\label{fig:diffus}
\end{figure}

The charge collected by the collection diode is the sum of the charges that can reach it directly and after reflection (figure~\ref{fig:diffus}). 
Because of the recombination of carriers, only a part of the charge can be collected by the diodes.
This is modelled by including an attenuation length $\lambda$.

The probability density $\Pi$ that a charge generated at Q reaches M is given by:
\begin{equation}
\Pi(\vec{R}) = \frac{d\Omega}{4\pi} \exp(-\frac{R}{\lambda}) \;  ,
\end{equation}
where d$\Omega$ is the solid angle from the track to M and $ R$ the distance MQ.

After the charge is distributed over individual pixels according to this equation, taking into account both direct and reflected pathways, a threshold is applied to determine whether the pixel fires.
Both the attenuation length and the threshold are not known a priori and are determined from the data, optimising the simulation of the cluster size and the lateral profiles.

Simulations were performed for the detector as-built, i.e. with its imperfections like non operating channels and sensors ("real detector"), and for the nominal detector with perfect components ("ideal detector"). Misalignment was taken into account, but the sensors were simulated with equal sensitivity and thickness ("perfectly calibrated").
The masking of excessively noisy pixels ("hot pixels", section~\ref{sec:tuning}) was not simulated as the fraction of pixels involved was less than one percent.

\section {Data Analysis} \label{sec:shower}

\subsection{Event Selection}
Due to the long integration time of 642 $\mu$s past-future protection is very important. 
All events with \emph{any} additional trigger within $\pm$642 $\mu$s were discarded. This is to prevent e.g. an event with a central particle (H \textsc{and} V scintillator) from being contaminated by a peripheral showering particle. Beam intensity was adjusted to retain on average 70\% of the events after past-future protection.
Furthermore, runs were inspected for stability, especially of the digital output channels. It happened that a channel spontaneously stopped working during a run. 
Although the analysis software was designed to cope with partially not working sensors, (part of) such a run was discarded in order to prevent normalisation problems.
In view of the limited transverse size of the detector, less than four Moliere radii, all analyses involving showers were done only for central ($|x| \le 10$ mm \textsc{and} $|y| \le$ 10 mm)  incoming particles. 
For tracking and alignment this condition was not imposed in order to have inclined tracks that would pass through different quadrants.
Instead, cuts were applied on the number of hits in the full detector and in the later layers to exclude interacting pions.

\subsection{Charged particle tracks} 

In this section the detector's response to charged non-showering particles is analysed. These  particles are supposed to leave straight tracks.
This is not completely true for the particles used, cosmic muons and charged pions of more than a few GeV. 
The cosmic muons scatter in the dense tungsten absorber (up to 0.02 rad) and a considerable fraction of the pions will have some interaction with the nuclei.
Therefore, all tracks that appear curved or multiple, were removed.
Although the notion "minimum-ionising" is not fully applicable, the term "MIP" will be used in the following for the sake of convenience.

From the events thus selected tracks are reconstructed according to the following procedure:
\begin{itemize}
\item Combine pairs of hits in different layers to tracklets.
\item Find the densest region in track parameter space.
\item Use the tracklet with the maximum number of neighbours as the proto track.
\item In each layer, assign hits within $r = 1 \, \mathrm{mm}$ to the proto track.
\item Accept only tracks with more than 24 hits in at least 12 different layers.
\item Find track parameters by $\chi^2$-minimisation of the distance of hits from a straight line.
\end{itemize} 
When performing analysis of a given layer, the track properties are calculated from all layers excluding the one in question.

\subsubsection{Alignment} \label{sec:align}
From the way the detector is assembled maximum deviations are expected of 0.2~mm in $ x, y, z$, 5~mrad in the in-plane angle $\phi$ and 2~mrad for the other angles.
For the reconstruction of tracks and showers, the accuracy in $z$ and the out-of-plane angles is sufficient and nominal values are assumed.
The remaining parameters ($x,y,\phi$ per sensor) need to be determined more accurately and for this cosmic muons are used.
Their tracks have a wide angular range and can thus connect the sensors in the different quadrants.

As there is no absolute position information in the cosmics measurements, see section \ref{sec:cosmic}, one needs to define the coordinate system relative to the sensor pixel matrix.
The first (row, column) pixel of sensor $(l,q)=(0,0)$ is taken as the origin: $x,y,z = 0$. 
The direction of the $z$-axis is defined by setting $x,y=0$ for the first (row, column) pixel of sensor $(l,q)=(23,0)$.
This leaves us with $(96-1) \times 3 - 2 = 283$ degrees of freedom, which are determined in a minimisation procedure. 
Because the parameters of the track used for determining the residuals will be influenced by the alignment itself, the whole procedure is iterated until the residuals are significantly smaller than $30 \, \mu \mathrm{m}$.

\begin{figure} [tbh]
\centering
\includegraphics[width=0.4\textwidth]{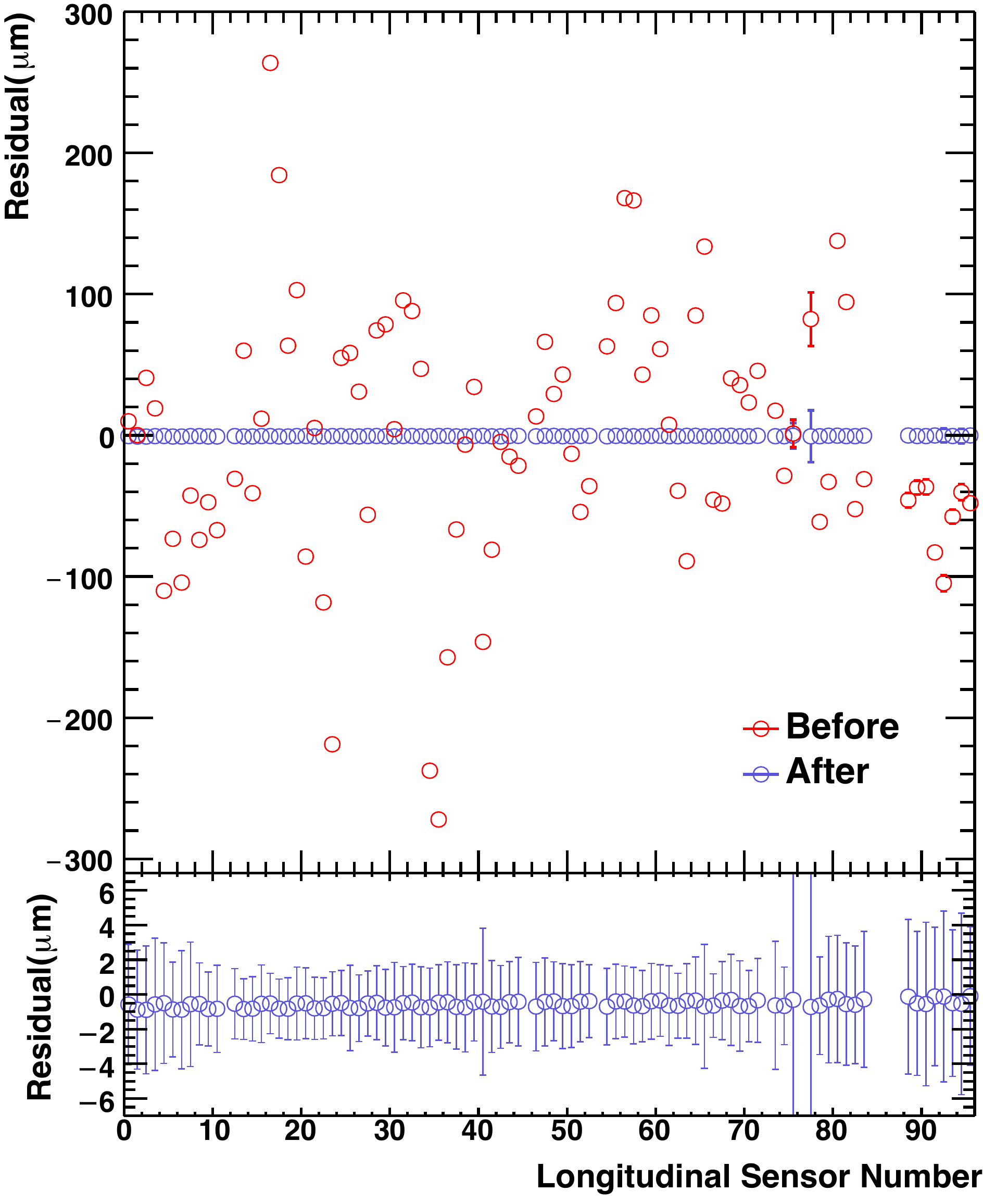}
\includegraphics[width=0.4\textwidth]{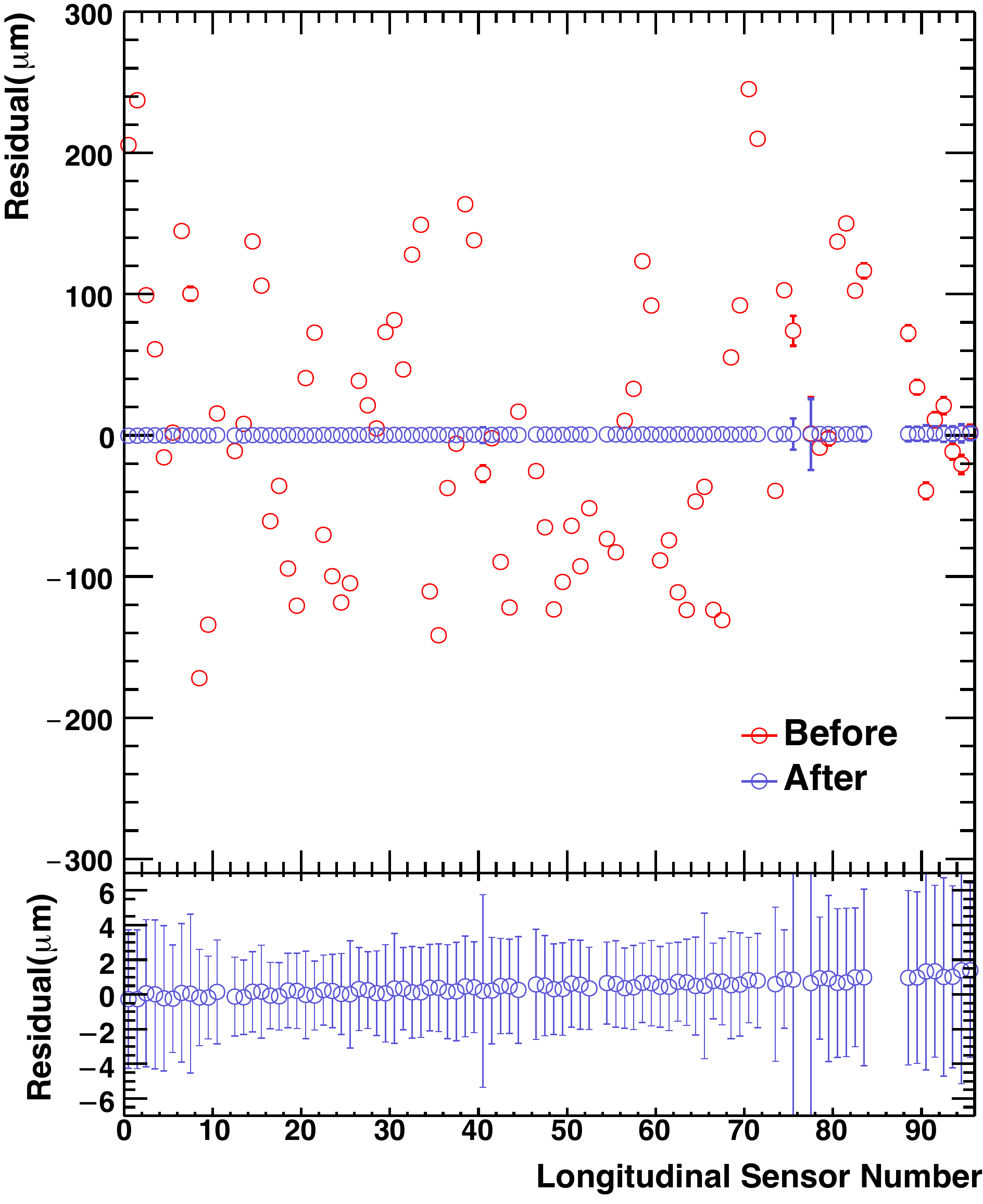}
\caption{Mean position residuals in $x$ (left) and $y$ (right) of all sensors as estimated from muon tracks. The red symbols show the residuals before alignment, the blue symbols after alignment (enlarged in the lower panels).
}
\label{fig:align}
\end{figure}

The alignment is checked by studying the residuals of a particular layer relative to the tracks. 
The residuals before and after alignment are shown in figure~\ref{fig:align}. The longitudinal sensor number is related to layer and position as $ n = 4\times l + q$.
While the sensors appear to have been positioned according to assembly accuracy, alignment improves this to $< 5 \, \mu \mathrm{m}$. 
Several cosmics data sets, taken before and after transport of the detector, gave the same results, proving the mechanical stability of the construction.

For the beam tests with hadronic or mixed beams a similar tracking algorithm is applied.
The detector thickness corresponds to approximately one hadronic interaction length, so many hadrons will not undergo any hadronic interaction and should leave a track similar to those of muons. Still, a significant fraction of the incoming particles might start a hadronic shower at some depth, and would thus deposit more energy and produce more hits in the detector. 
For the tracking analysis of these samples a maximum number of 1000~hits in the whole detector is required. Such a cut will also reject any electromagnetic shower present in mixed beams -- the resulting samples should thus be dominated by pion tracks.

These pion tracks were also used to determine the inclination angle, i.e. the angle between the beam axis and the detector axis.
For most of the measurements this small ($\approx 10$ mrad) inclination is not crucial. However, all analysis which depends on the exact position of the incoming particle trajectory with respect to a given layer will use a position estimate corrected for this inclination. This is particularly important for a detailed study of the lateral profiles.
The divergence of the beam ($\approx 1$ mrad) is neglected.

\subsubsection{Response to MIPs} \label{sec:mipresponse}

\begin{figure} [tbh]
\centering
\includegraphics[width=0.4\textwidth]{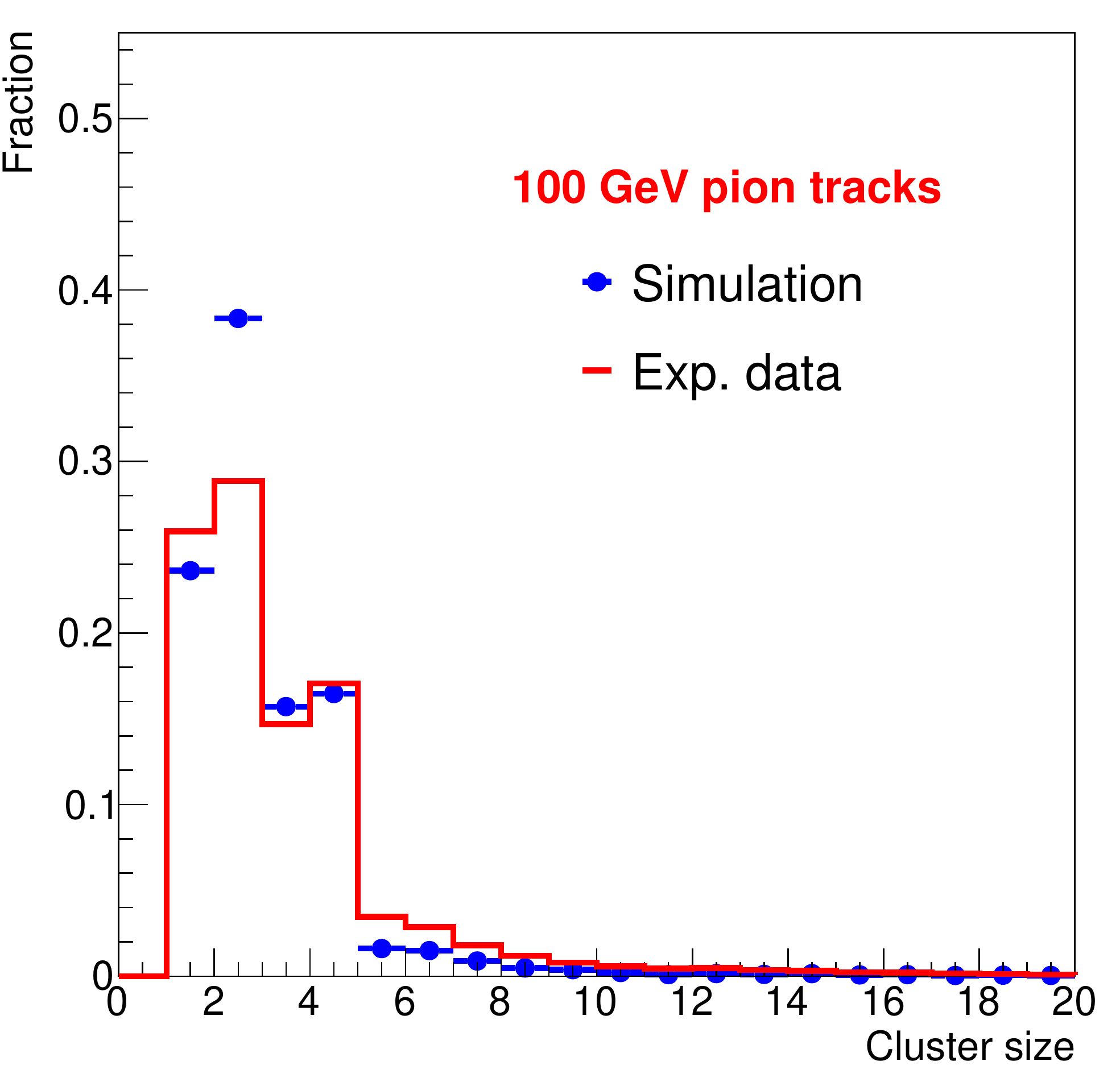}
\includegraphics[width=0.4\textwidth]{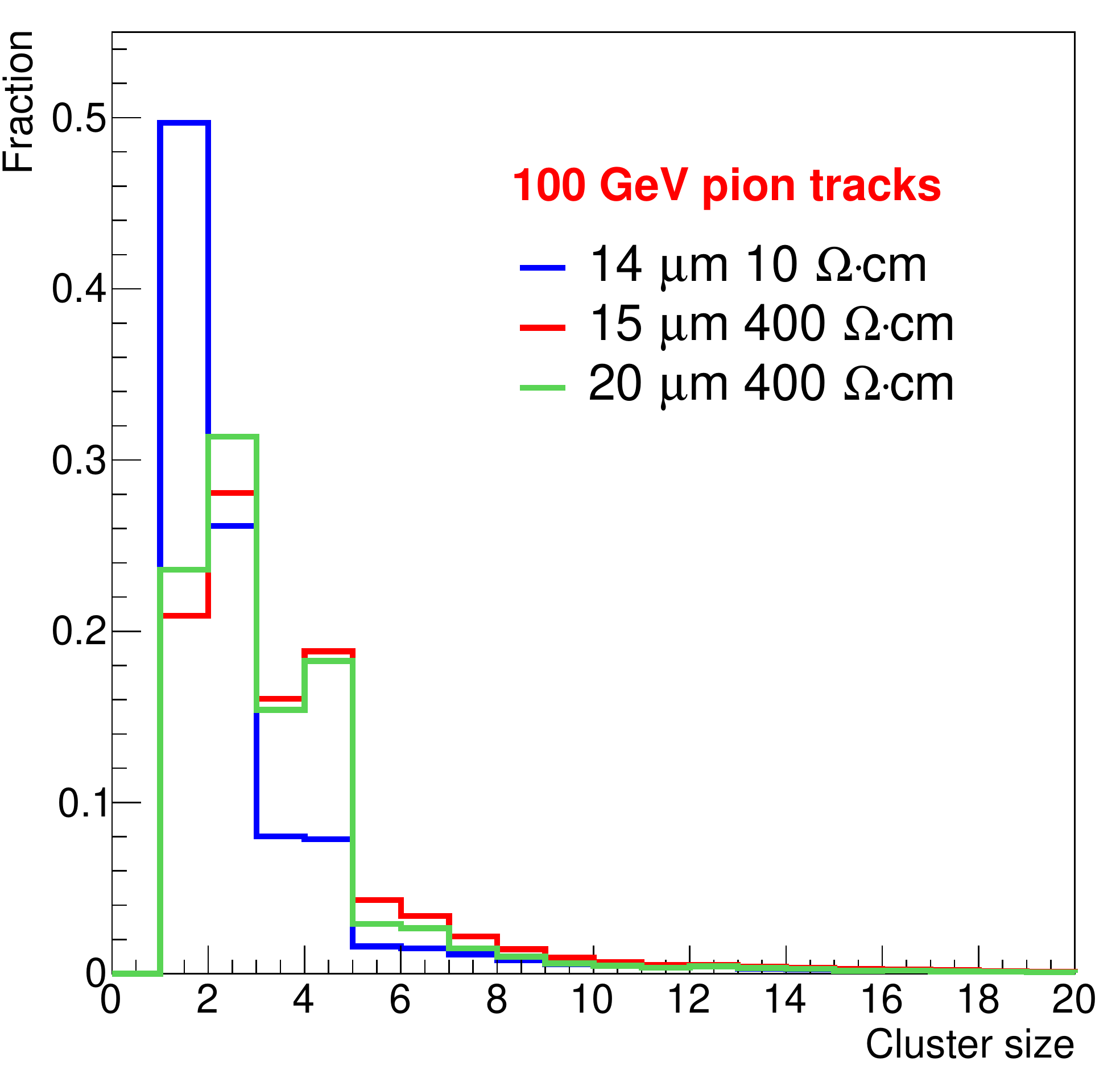}
\caption{Measured distributions of cluster sizes for 100 GeV/$c$ pion tracks: \textit{left} for all sensors compared to the result of Monte Carlo simulations, 
\textit{right} for the three types of epitaxial layer: 400~$\Omega$cm of 15 and 20~$\mu$m thickness, respectively,
and 10~$\Omega$cm, 14~$\mu$m.
}
\label{fig:mipcluster}
\end{figure}

The tracks from muons and pions can be used to study the sensor response to minimum-ionising particles (MIPs). To study the response in a given sensor, tracks have been selected from all sensors excluding the one in question. Then associated clusters are defined as contiguous areas of hit pixels in the sensor within a distance of $r = 1 \, \mathrm{mm}$ from the track.
In a fraction of the cases no hit pixels are found due to a too small signal (dead pixels are removed from the procedure).
From this one can derive the detection efficiency of each sensor. Table~\ref{tab:eff} shows the efficiency for the three types of sensors.
 Figure~\ref{fig:mipcluster} left shows the distribution of the corresponding cluster sizes of the tracks of pions. 
The cases with no hits (cluster size 0) are excluded.
The average cluster size then is $\approx 3$. The results from Monte Carlo simulations are also included in the figure.
The discrepancy between data and simulation may be due to possible inhomogeneity within a sensor.

\begin{table}[h]
\caption{Detection efficiency for pions of 100 GeV/$c$.}
\begin{center}
\begin{tabular}{|c|c|}
\hline
sensor type & efficiency from data \\ \hline
14 $\mu$m 10 $\Omega$cm & 0.85 \\ \hline
15 $\mu$m 400 $\Omega$cm & 0.98 \\ \hline
20 $\mu$m 400 $\Omega$cm & 0.98 \\ \hline
\end{tabular}
\end{center}
\label{tab:eff}
\end{table}

Another outcome of this analysis is the influence of the sensor thickness on the cluster size.
One would expect less charge collected in the low resistivity epitaxial layer due to recombination and one would also expect a dependence on its thickness.
How this translates into cluster size depends on the diffusion, which may be different.
Figure~\ref{fig:mipcluster} right shows the measured cluster size histograms for the three different types of epitaxial layer.
While the low-resistance material produces smaller clusters, the dependence on epitaxial layer thickness is small.

\subsection{Electron showers}

In this section events are considered which are caused by an incoming electron in the central region of the detector.
At some beam momenta, contamination with fully developed pionic showers cannot be excluded, as no external particle identification was present.
This contamination is estimated to be less than 1\%.

\subsubsection{Position determination} \label{sec:posdet}
The position of a track can be very precisely determined as discussed in the previous section~\ref{sec:align}. This is different for showering particles. In general, the exact position of incidence of the charged particle is derived from layer 0 of the detector. 
In this layer the shower has not yet started and the signal may amount to only a few (or even a single) pixel(s), risking contamination by noise hits in the corresponding sensors. The selection procedure is as follows:
\begin{itemize}
\item A first approximation to the shower position is obtained from the lateral centre of gravity of the hit distributions in layers 3 and 4. There are several reasons for taking these layers: 1) at the lowest energy (2~GeV) the shower maximum is expected in layer 4, ensuring sufficient statistics, 2) the lateral size of the showers is still small, 3) these layers are among the most homogeneous in sensitivity.
\item Clusters of hits are identified within a radius of $ r = 1 \, \mathrm{mm}$ around the position in layer 0 as estimated from layers 3 and 4.
\item Events where more than one cluster is found in layer 0 within $ r = 1 \, \mathrm{mm}$ or where the cluster size is larger than 12 pixels are rejected.
\item For the accepted events the centre of gravity $( x_\mathrm{N}, y_\mathrm{N})$ of the correlated cluster in layer 0 is taken as the nominal position of the shower centre.
\end{itemize}
The position of the incoming charged particle obtained this way is the one to be used in the test beam analysis\footnote{Note that, while information from layers 3 and 4 is used to identify the appropriate cluster in layer 0, the nominal position is calculated from the pixel distribution in layer 0 only.}.

\begin{figure} [tbh]
\centering
\includegraphics[width=0.5\textwidth]{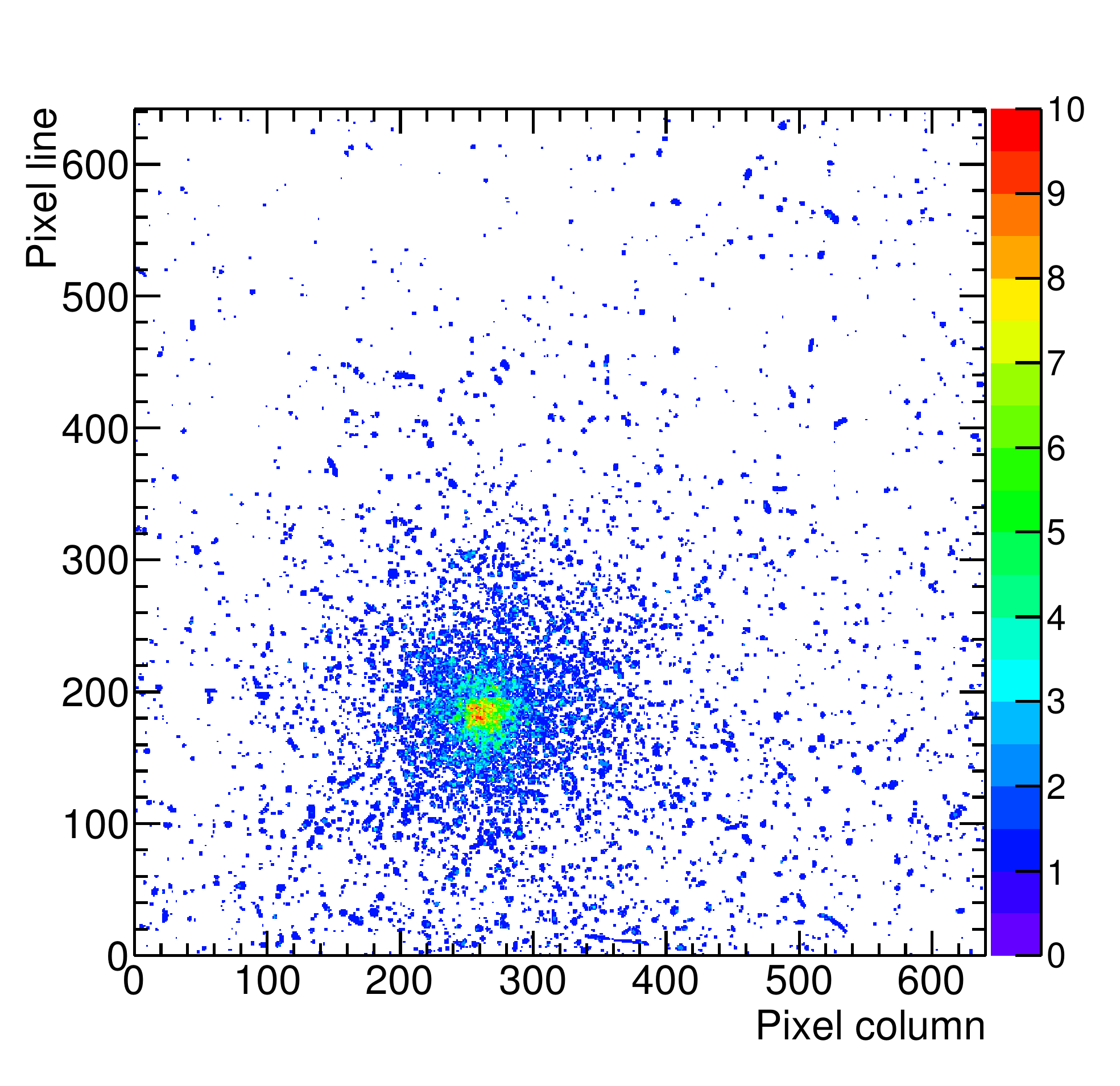}
\caption{All hits in quadrant $q=0$ of a single 100 GeV electron according to equation \protect \ref{eq:position1}.
}
\label{fig:hitmapAB}
\end{figure}

Note that this procedure can only be applied for charged particles (electrons) and not for photons.
A realistic estimate of the shower position would make use of all layers except layer 0 and should be applicable for both electrons and photons.
The procedure is as follows. One defines a rectangular region with line and column numbers \{$s,u$\} around $( x_\mathrm{N}, y_\mathrm{N})$ and sums the hits in all layers:
\begin{equation}
h(i,j) = \sum_{l=1}^{23}  w(i,j,l)  \; \mathrm{  for  }\; \{i,j\} \in \{s,u\} \; ,
\label{eq:position1}
\end{equation}
where $w(i,j,l)$ is the value (0 or 1) of pixel (line $i$, column $j$, layer $l$) for good sensors.
Figure~\ref{fig:hitmapAB} shows $h(i,j)$ for one electron event  of 100 GeV where the region $\{s,u\}$ corresponds to quadrant $q=0$.

Then the position in $x$ is obtained by
\begin{equation}
 x_\mathrm{S} = \frac {\sum_{\{i,j\} \in \{s,u\}} \left[ h(i,j) \cdot \theta \left( h(i,j)-d \right) \right] ^p \cdot  x(i,j)} 
{\sum_{\{i,j\} \in \{s,u\}} \left[ h(i,j) \cdot \theta \left( h(i,j)-d \right) \right] ^p} \;  ,
\label{eq:position2}
\end{equation}
and likewise for $y_\mathrm{S}$.
A threshold $d$ is implemented here via the Heaviside step function $\theta$.
The values for the parameters $p=2$ and $d$ are obtained from optimisation.

\subsubsection{Lateral hit distributions} \label{sec:rawlateral}

The lateral hit distributions of showers as a function of the distance $r$ from the nominal shower centre:
\begin{equation}
r \equiv \sqrt{(x-x_\mathrm{N})^2 + (y-y_\mathrm{N})^2}
\label{eq:distance}
\end{equation}
were obtained by counting the number of hits in a given layer in rings around this centre 
and normalising to the active area in the ring for each incoming particle. The hit density per unit area for sensor $q$ in layer $l$ is given by:
\begin{equation}
\nu_{l,q}(r) =  
\frac{\Delta N_{\mathrm{hit}}(r,\Delta r; l; q)-\Delta N_{\mathrm{noise}}(r,\Delta r; l; q)}{\Delta N_{\mathrm{pixel}}(r,\Delta r; l; q) \cdot (30 \, \mu \mathrm{m})^2}  \approx \frac{1}{2 \pi r} \frac{dN^{(l)}}{dr}\;  ,
\label{eq:lateral}
\end{equation}
where $\Delta N_{\mathrm{hit}}$ is the number of hit pixels in a ring of width $\Delta r$ at radius $r$.
 Likewise, $\Delta N_{\mathrm{noise}}$ is the number of noise pixels expected in sensor $(l,q)$ based on the pedestal runs.
 $\Delta N_{\mathrm{pixel}}$ represents the total number of live pixels in this ring.
Due to the square shape of the pixels the true shape of a ring is only approximated, but this is taken into account both in counting the hits and normalising the area.

In this way, one automatically corrects for dead areas or masked pixels in the sensors, as well as for the gap and overlap areas. 
In the following angle brackets $\langle\rangle$ indicate averages over events.

\begin{figure}  [tbh]
\centering
\includegraphics[width=0.4\textwidth]{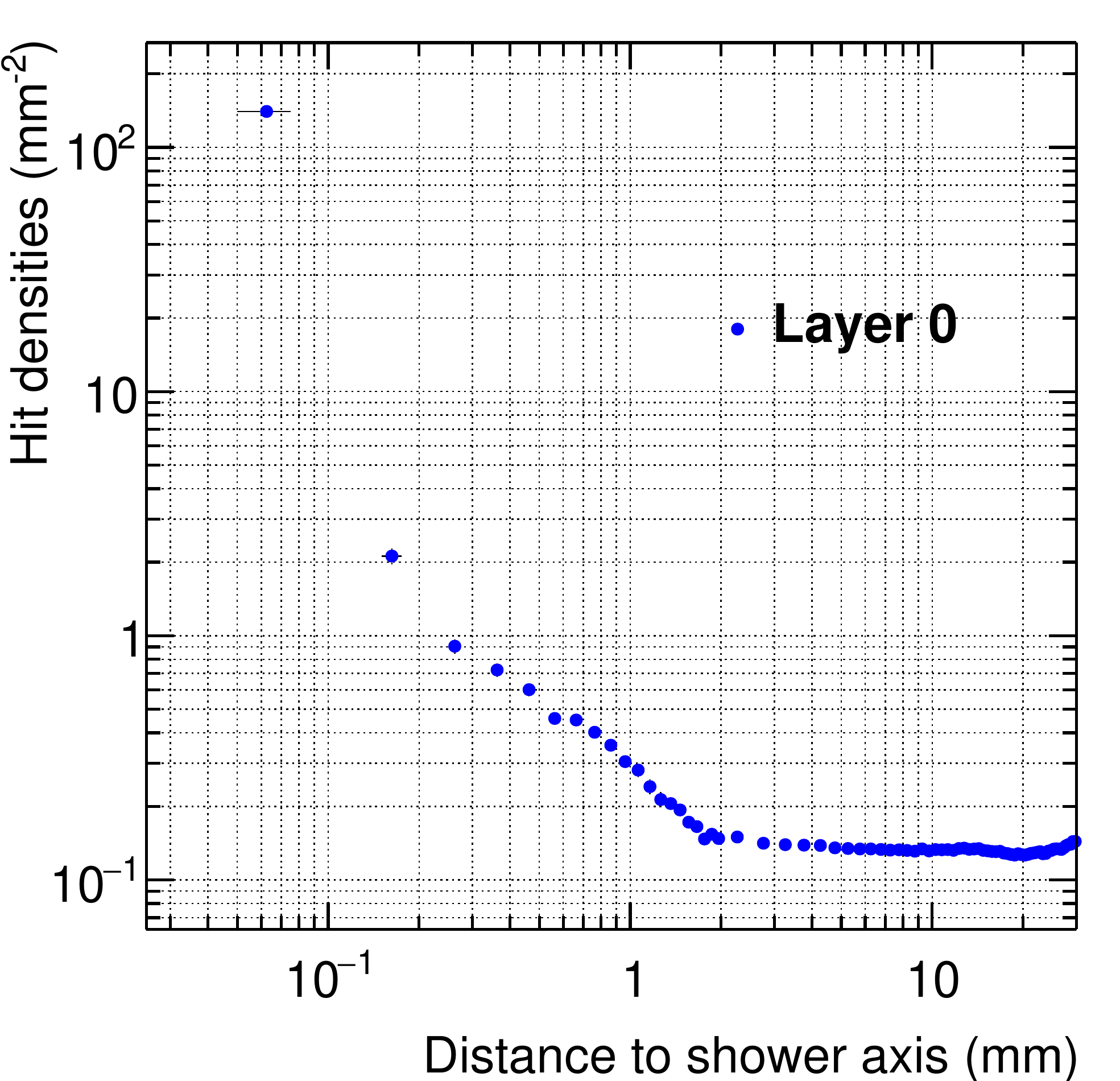}
\caption{Hit density (raw data, noise not subtracted) per area as a function of distance from the shower centre for electrons of $100$~GeV as measured with layer 0, which acts as a charged particle detector. Note the double logarithmic scale.}
\label{fig:lateral0}
\end{figure}

 We have checked the behaviour of layer 0 used for the determination of the shower centre. Figure~\ref{fig:lateral0} shows the hit density distribution as a function of distance $r$ from the shower centre using equation~\ref{eq:lateral}, but without noise subtraction.
 At larger radii $r$ one observes a constant hit density of $\approx 0.1 \, \mathrm{mm}^{-2}$ as expected from the noise level in this layer. In addition, there is a peak of $10^2\, \mathrm{mm}^{-2}$ for $r < 0.1$mm -- this corresponds to an average number of hit pixels per incoming electron of $\approx 3$,
which is in line with the measured cluster size distribution. 
The width of the peak reflects the width of the cluster used for the position of the nominal shower centre -- its sharpness enables a high accuracy of the reference position measurement.
The hits in the range $r$ = 0.2 \textellipsis 2 mm are probably due to early conversions and backscattering.

\subsection{Sensor Calibration} \label{sec:calib}

The sensitivity of individual sensors to charged particles and thus their contribution to the calorimetric energy measurement may vary.
 The prototype makes use of three types of sensor chips with different properties (resistivity and thickness of the sensitive layer), resulting in different responses as was shown in figure~\ref{fig:mipcluster}. 
So a relative calibration due to different sensor sensitivities is necessary.
The uniformity for each sensor was optimised on the basis of the noise, but there could be a non-uniform sensitivity. This internal sensitivity variation will be ignored here and only  inter-sensor calibration will be attempted.

\begin{figure}  [tbh]
\centering
\includegraphics[width=0.4\textwidth, angle=90]{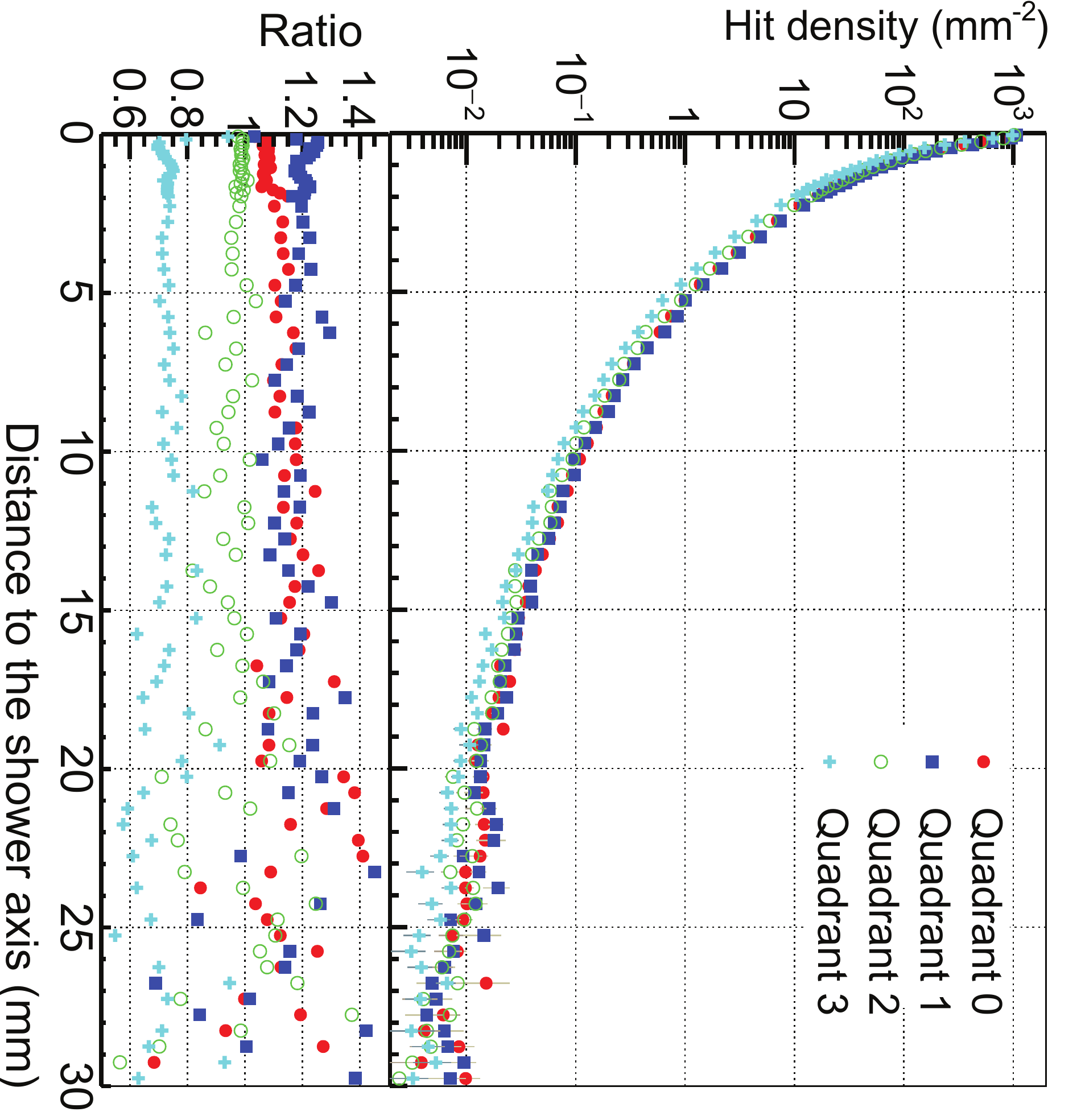}
\includegraphics[width=0.4\textwidth, angle=90]{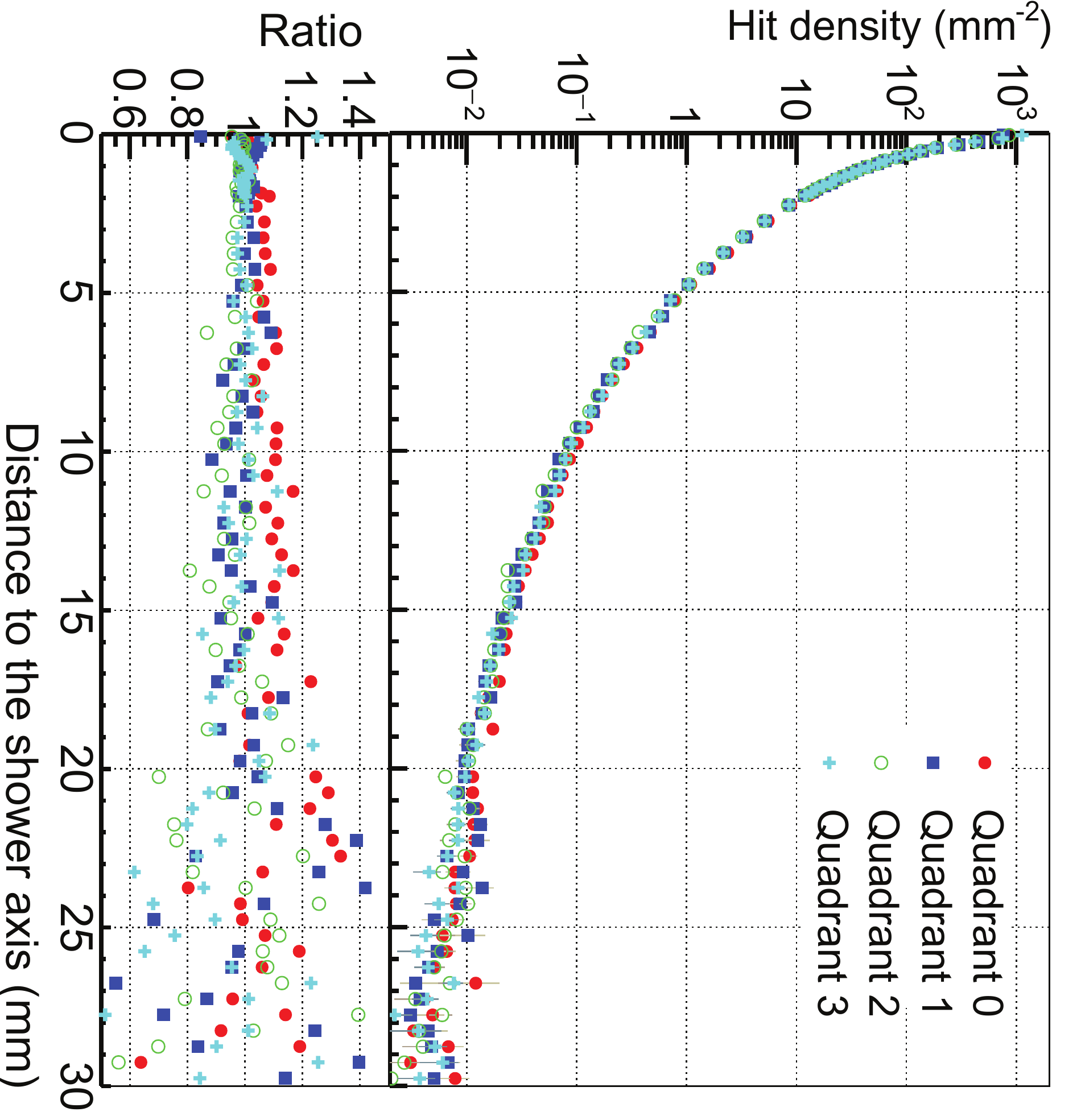}
\caption{Average hit density in layer 4 for electrons of 50~GeV of individual sensors (upper panels) and normalised to their average in this layer (lower panels) before (left)  and after (right) calibration. Note that the calibration procedure uses only data with $r \le 18$~mm.}
\label{fig:lat-chips}
\end{figure}

In a first calibration step the average lateral distributions $\langle \nu_{l,q}(r) \rangle$ from equation \ref{eq:lateral} will be compared for the different sensors $q =0,1,2,3$ independently in the same layer $l$, where the behaviour of the lateral distributions should be similar. 
As an example, figure~\ref{fig:lat-chips} (left)) shows $\langle \nu_{4,q}(r) \rangle$ in the upper panel and the ratios
\begin{equation}
V_{l,q}(r) = \frac{4 \langle \nu_{l,q}(r) \rangle}{\sum_0^3 \langle \nu_{l,q}(r) \rangle} \; 
\end{equation}
for $l=4$ in the lower panel.
Clearly, there is a difference in sensitivity of these four sensors, with $q = 1$ the most sensitive.

The cumulative hit densities per sensor are defined as
\begin{equation}
 M_{l,q} = \int_{0}^{R} 2 \pi r \langle \nu_{l,q}(r) \rangle dr \; ,
\end{equation} 
where the integration proceeds until $ R$ = 18 mm, and for the full layer:
\begin{equation}
M_l = \sum_0^3  M_{l,q} /4.
\end{equation} 

The per sensor calibration factor is then obtained by
\begin{equation}
  c_{l,q} =\frac {M_l}{M_{l,q}} \; .
\end{equation}

This way, the sensors within each layer are calibrated with respect to each other, as can be seen for layer 4 in figure~\ref{fig:lat-chips} (right).

As a second step, the longitudinal hit distribution $G_l$ is calculated:
\begin{equation}
G_l = \sum_{q=0}^3  c_{l,q} M_{l,q} \; .
\label{eq:latintegral}
\end{equation}

\begin{figure} [tbh]
\centering
\includegraphics[width=0.5\textwidth]{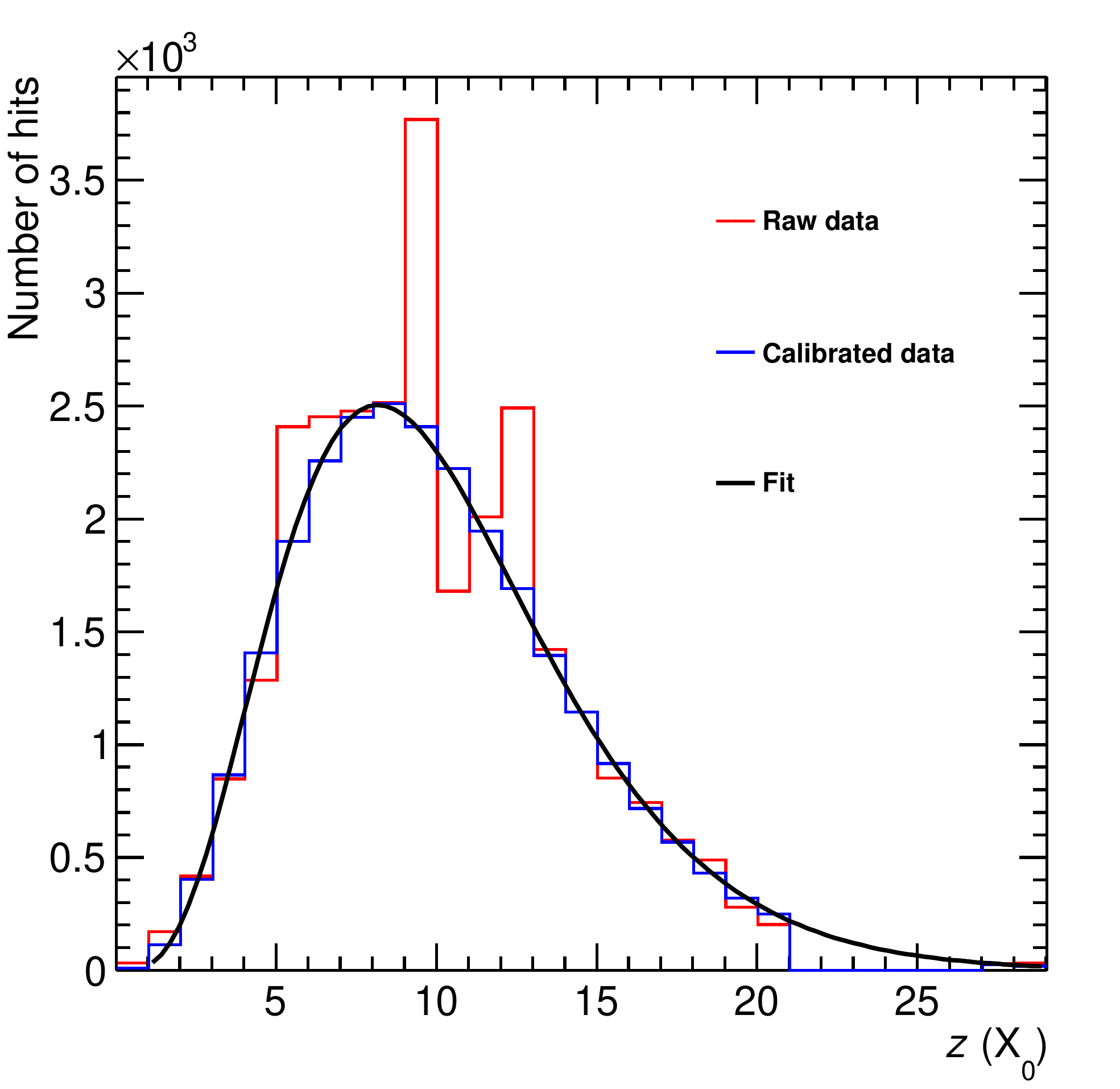}
\caption{Total number of hits per layer as a function of $z$-position for 100~GeV electrons: raw data (red) and calibrated data (blue). The black line shows the function of eq. \protect \ref{eq:longprof}. Note that the depth range $ z = 21 X_0 ... 26 X_0$ is not instrumented.}
\label{fig:long0}
\end{figure}

This is shown in figure~\ref{fig:long0}, red curve. 
Overall, the distribution shows the expected development of a shower; however one can also clearly observe layer-by-layer variations.
Here, one cannot use any simple symmetry argument to perform a layer-by-layer calibration and has to rely on a theoretical shape.
 
The calibration procedure first fits the function \cite{pdg}
\begin{equation}
 N(t) = N_0 b  \frac{(bt)^{a-1}\exp(-bt)}{\Gamma (a)}
\label{eq:longprof}
\end{equation}
to $G_l$ to obtain the parameters $N_0$, $a$ and $b$. $N$ is the number of hits and $t = z /X_0$ is a discrete function of the layer number $l$.
 The relative calibration factors of different layers $k_l$ are then chosen such to reproduce this function (blue curve in figure~\ref{fig:long0}).
Obviously, such a procedure omits any detailed information on the average longitudinal distribution of electron showers, and it relies on the assumption that the function (\ref{eq:longprof}) describes the data well.
As a check, a miscalibration was implemented in the simulations and it was found that this procedure does not introduce a significant bias in the global features of the longitudinal distribution, like e.g. the longitudinal position of the shower maximum.

Finally, the calibrated hit density is obtained by:
\begin{equation}
\rho_{l}(r) = \sum_{q=0}^3 c_{l,q} k_l \nu_{l,q} \;  .
\label{eq:latprof}
\end{equation}
In cases where $\Delta N_{\mathrm{hit}}(r,\Delta r; l; q) = 0$ because the shower occurs in a dead area, the hit density is interpolated according to:
\begin{equation}
 	\rho_l(r) = \frac{1}{2}[\rho_{l-1}(r)+\rho_{l+1}(r)] \; .
 \label{f:longitudinalDensities}
\end{equation}
 The calibrated number of hits in a layer is given by
\begin{equation}
 P_{l} = \int_{0}^{R} 2 \pi r  \rho_{l}(r)  dr \; ,
\label{eq:layerresponse}
\end{equation} 
where the integration proceeds until $ R$ = 22  mm or the edge of the sensors is reached.

\subsubsection{Calibration using MIPs}

One would expect that the cluster size of MIPs obtained in section \ref{sec:mipresponse} be related to a sensor's sensitivity and that one could use this in a way analogous to the MIP calibration of a classical calorimeter. 
Various attempts to calibrate the detector in such a way failed, see \cite{martijn, chunhui}.
Also in the present analysis only a weak correlation was seen between the average cluster size and the calibration factors derived above.

The most likely explanation is related to the assumption that the response to a shower is the sum of the responses of minimum-ionising particles.
One reason why this is not correct is that not all shower particles are MIPs.
Moreover, they may travel at small angles with respect to the sensor, leading to high density features.
Obviously, these are not signatures of MIPs.
Another reason is that the cluster size (number of pixels above threshold) of overlapping clusters can range from slightly above the size of a single cluster to even more than the sum of the two clusters.
Last but not least, the relation between the size of a cluster and the energy deposited by the particle is not linear due to the interplay between charge diffusion and discriminator threshold setting.

\section{Results} \label{sec:results}

\subsection{Lateral Shower Profiles}

\begin{figure}  [tbh]
\centering
\includegraphics[width=0.45\textwidth]{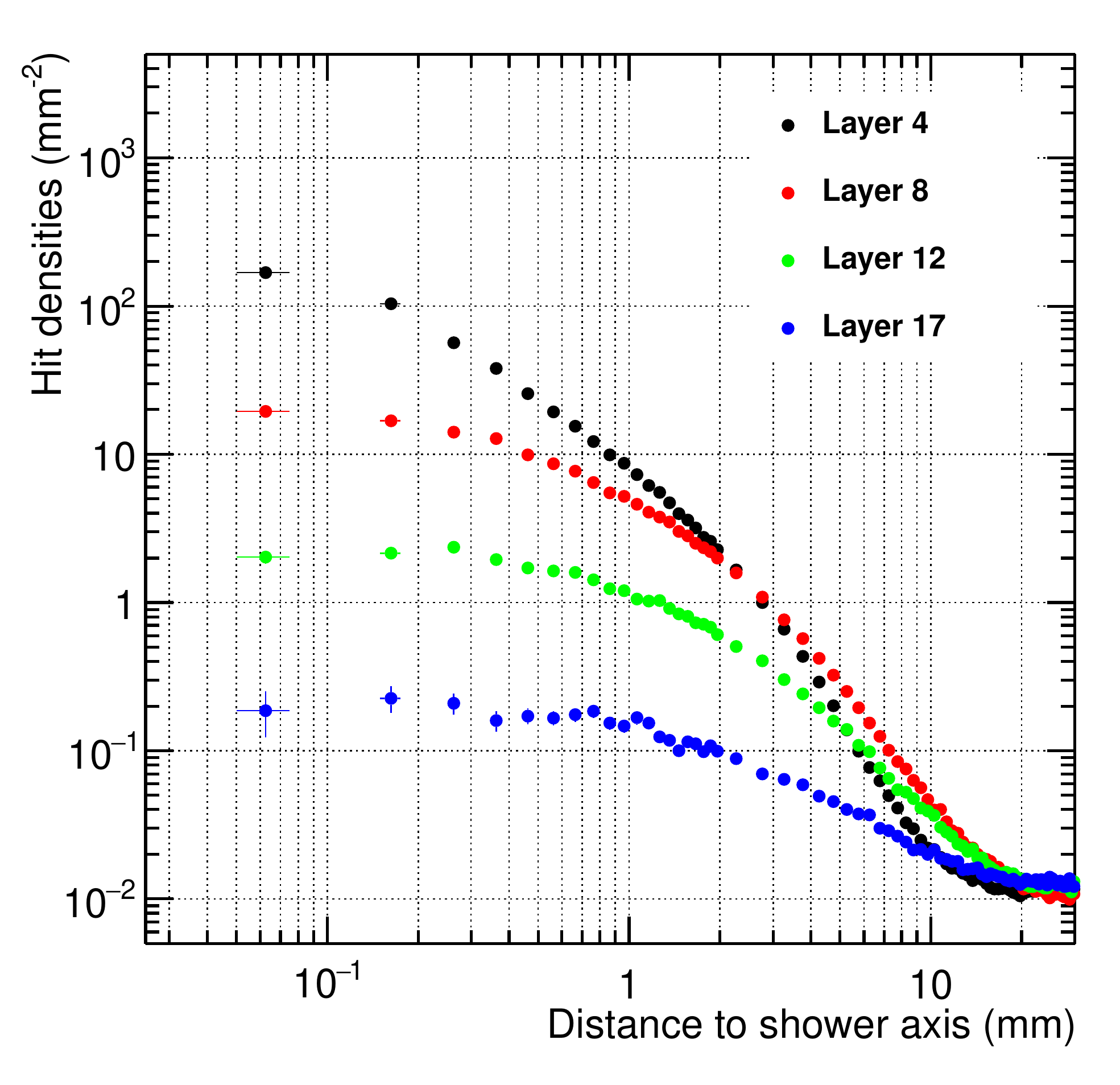}
\includegraphics[width=0.45\textwidth]{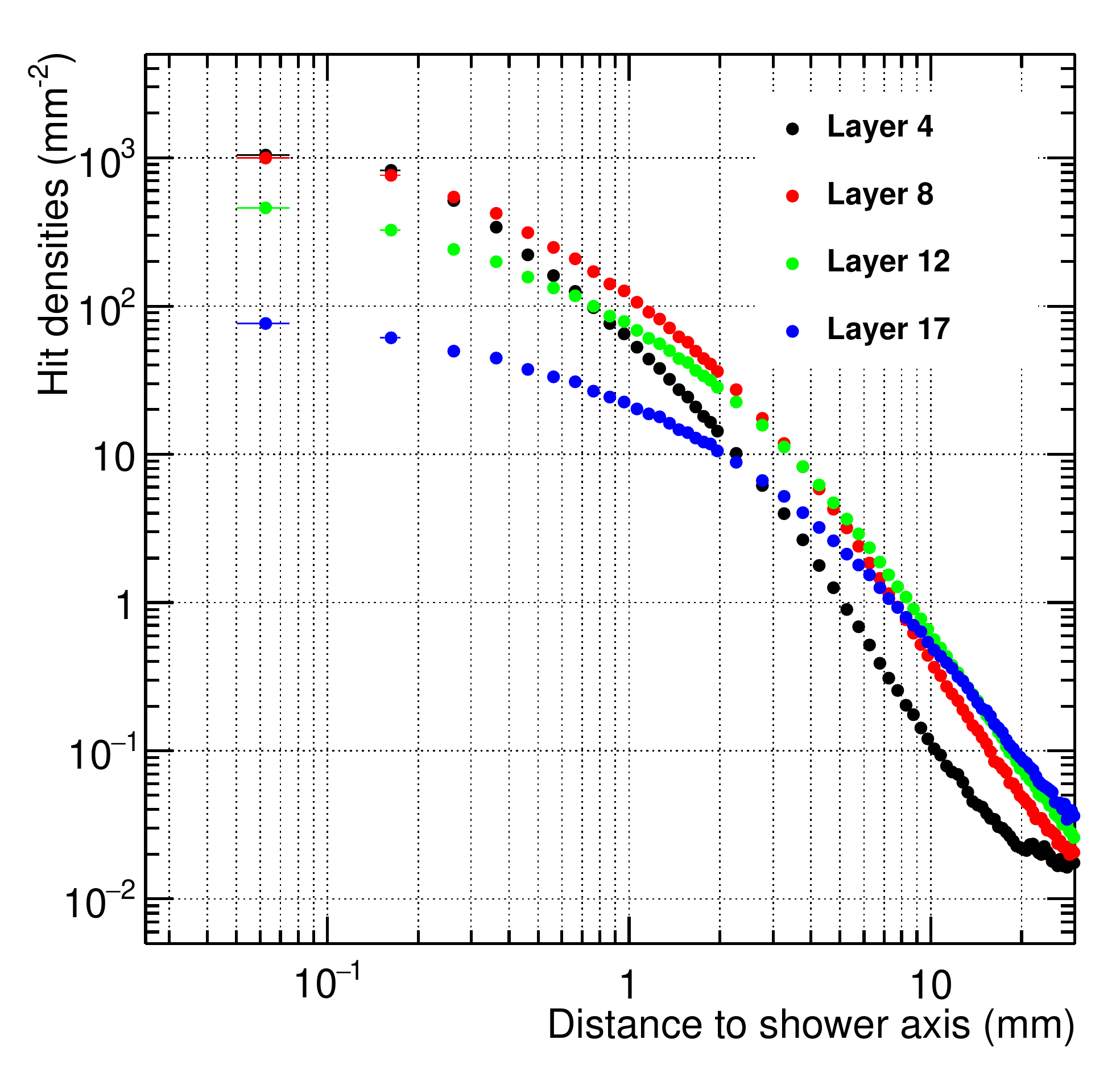}
\caption{Lateral shower profile for electrons of 5.4~GeV (left) and 100~GeV (right) for selected layers (radiation depth). Note the double logarithmic scale.}
\label{fig:lateral}
\end{figure}

Figure~\ref{fig:lateral} shows lateral profiles (calibrated hit densities $\langle\rho_l (r)\rangle$ with noise added back\footnote{After noise subtraction, the values of the number of hits in the tail of the distributions fluctuate around zero, in particular for low energy. The noise has been added back, because this allows displaying the distributions on a logarithmic scale.}, averaged over all events) for selected layers, where the layer number corresponds approximately to the number of radiation lengths in the detector.

 For 5.4 GeV layer 4 shows a very steep profile with a maximum of $  10^2 \, \mathrm{hits/mm}^2$, which falls by almost two orders of magnitude over a radial distance of less than 2~mm. It decreases further to level out at the noise level at $r \approx 20 \, \mathrm{mm}$.
 Deeper in the calorimeter the distributions get successively flatter behind the shower maximum, and contain fewer hits.
For 100 GeV electrons the profile in layers 4 and 8 remains very steep and the maximum at $r < 0.1 ~$mm stays about constant. 
The maximum density observed here is close to the saturation value of $ 1.1 \cdot 10^3 \, \mathrm{hits/mm}^2$, which implies that non-linearity effects start to play a role from this energy onwards.
Layers 12 and 17, which are beyond the shower maximum, show an increasing width and a decreasing density.
Even at 30~mm ($\approx2.5 R_\mathrm{M}$) from the core the density is still a factor 2 above the noise level, which shows that a small $\mathcal{O}(10^{-4}$) lateral leakage occurs.

\begin{figure}  [tbh]
\centering
\includegraphics[width=\textwidth]{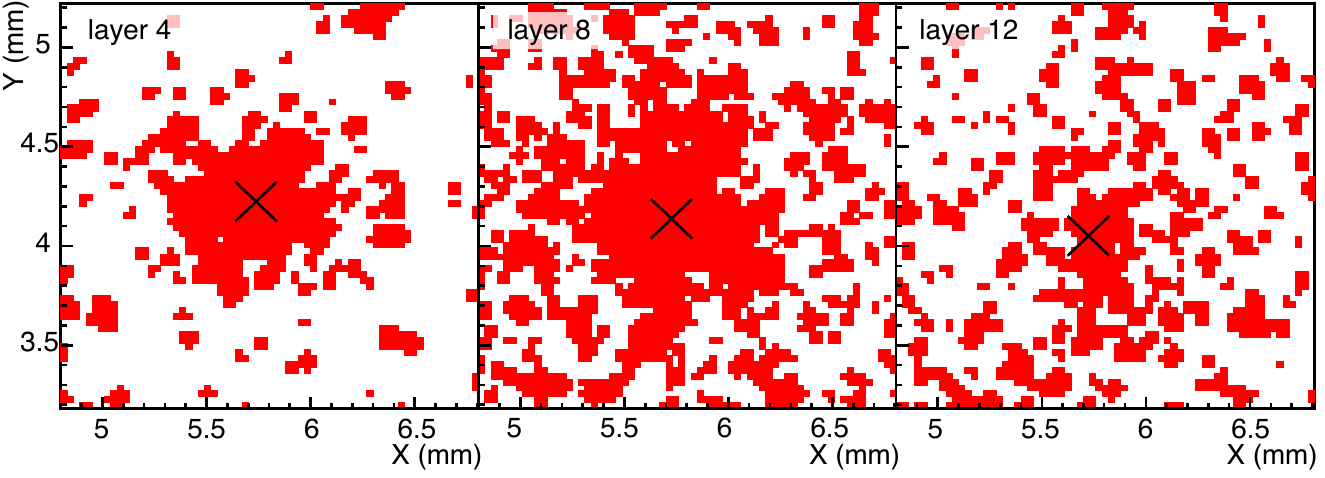}
\caption{All hits produced around the shower centre (cross) by a single electron of 244 GeV in layer 4 (left), layer 8 (centre) and layer 12 (right).}
\label{fig:coredensity}
\end{figure}

To further illustrate possible saturation, figure~\ref{fig:coredensity} shows a close-up of the core of a single shower in three layers around shower maximum, 
where the hit density is expected to be maximal, at the highest measured energy of 244~GeV.
Clusters are much larger than the typical cluster size shown in figure~\ref{fig:mipcluster}, indicating overlapping underlying charge distributions.
Clearly there is no longer a linear relation between the number of particles and the number of hits or clusters.
Full occupancy only occurs in the central millimetre.

\begin{figure}[tbh]
\centering
\includegraphics[width=0.4\textwidth]{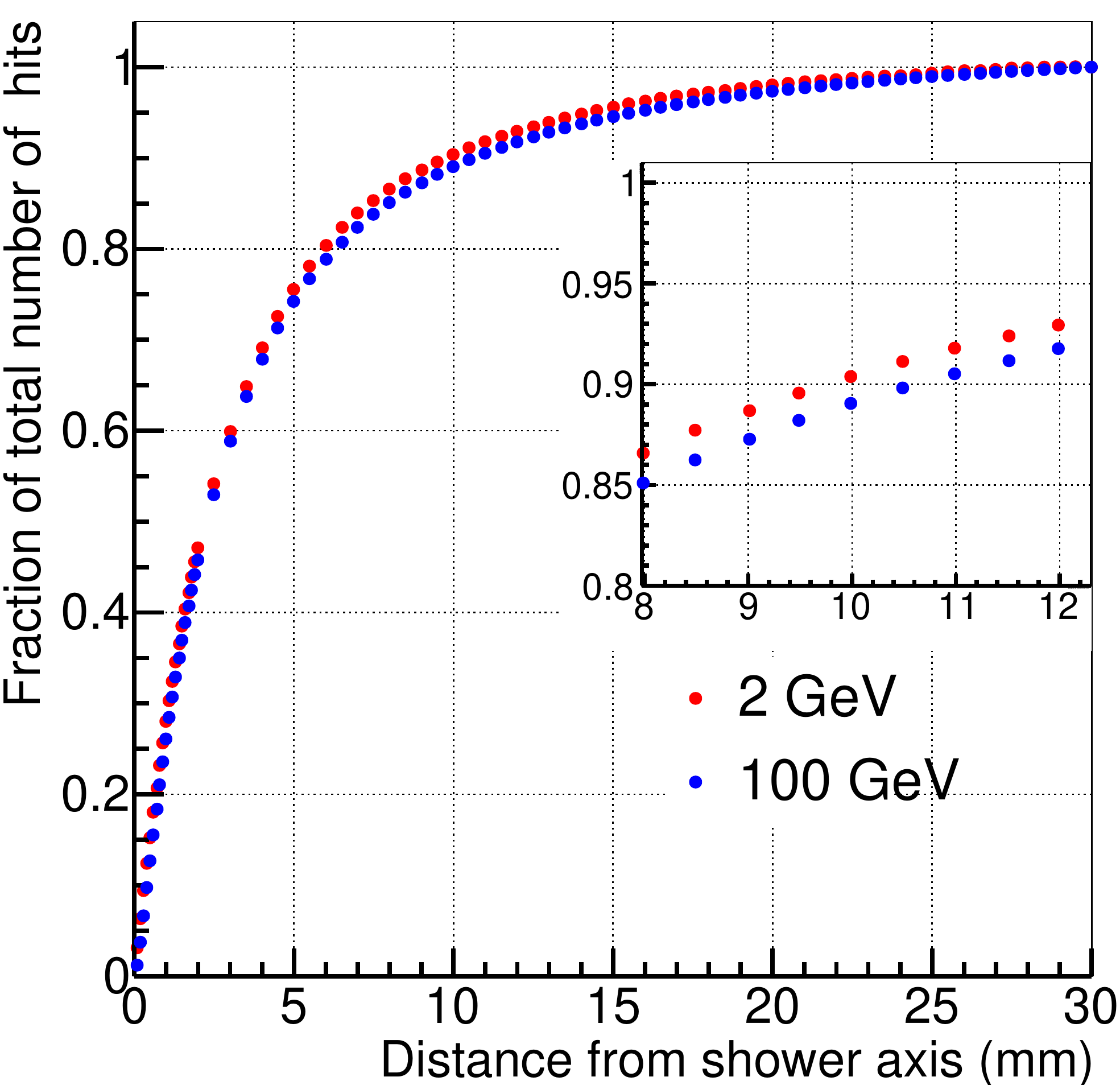}
\caption{Cumulative distribution of hits as a function of  distance from the shower centre for electrons of 2~GeV (red) and 100 GeV (blue) integrated over all layers, then normalised.}
\label{fig:molradmeas}
\end{figure}

Integration of the lateral profiles gives cumulative distributions of hits within a certain radius.
By summing these  for all layers and normalising them to the value at the largest radius obtainable (figure~\ref{fig:molradmeas}),
one can estimate the Moli\`ere radius, i.e. the 90th percentile.
From this figure one reads $R_\mathrm{M} = 9.9 \pm 0.5 \, \mathrm{mm}$ at 2 GeV,
 in agreement with the value of 10.7 mm derived from the material composition of the detector design.
One can see that the shape of these distributions and thus the Moli\`ere radii obtained show some energy dependence.
 The value of $R_\mathrm{M}$ extracted is slightly larger at the higher energy.

\subsection{Shower Position Resolution}

\begin{figure} [tbh]
\centering
\includegraphics[width=0.4\textwidth]{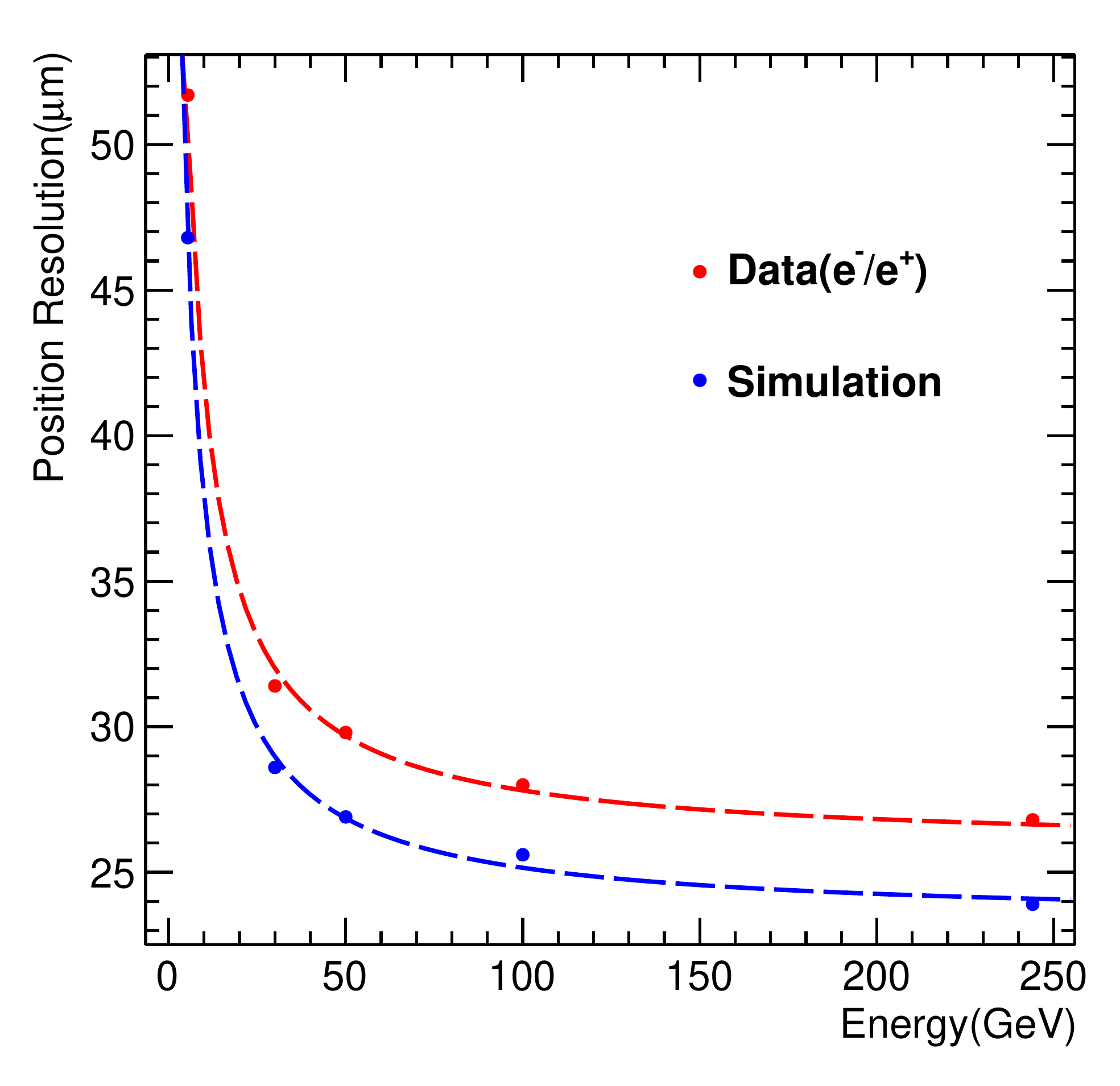}
\caption{Position resolution in $x$ for electron showers in the energy range of 5.4 to 244 GeV. Note: vertical scale does not start at 0.}
\label{fig:PosRecEn}
\end{figure}

Using the method outlined in section \ref{sec:posdet} the shower position $( x_\mathrm{S}, y_\mathrm{S})$ was determined at various energies.
The distribution of the event-by-event differences between this shower position and the nominal position  $( x_\mathrm{S} - x_\mathrm{N}, x_\mathrm{S} - x_\mathrm{N})$ has been used to obtain the position resolution.
The position resolution $\sigma_x$ as a function of energy is given in figure~\ref{fig:PosRecEn} for both experiment and simulation.
The experimental data can be fitted with
\begin{equation}
 \sigma_x = f  \oplus  \frac{g}{\surd E } \; ,
\end{equation}
with $f = 25.8 \pm 0.3\; \mu \mathrm{m}$ and $g = 104 \pm 1 \; \mu \mathrm{m}\sqrt {\mathrm{GeV}}$.
The simulation is based on the "real detector" (see section \ref{sec:simul}) and has parameters
$f = 23.3 \pm 0.2 \; \mu \mathrm{m}$ and $g = 94 \pm 1 \; \mu \mathrm{m}\sqrt{\mathrm{GeV}}$.
The results of the measurements are marginally worse compared to the simulation.
This is expected, because the simulation does not include all effects present in the experimental data which could deteriorate the position resolution, like the small, but finite angular divergence of the beam.
Note, that the very high granularity allows an extremely good position resolution, more than two orders of magnitude smaller than the Moli\`ere radius.

\subsection{Energy Linearity and Resolution}

The calibrated number of hits in the full calorimeter is obtained per event from:
\begin{equation}
K = \sum_{l=1}^{23}  P_l \; .
\label{eq:response}
\end{equation}

\begin{figure}  [h]
\centering
\hspace{-2 cm}
\includegraphics[width=0.5\textwidth]{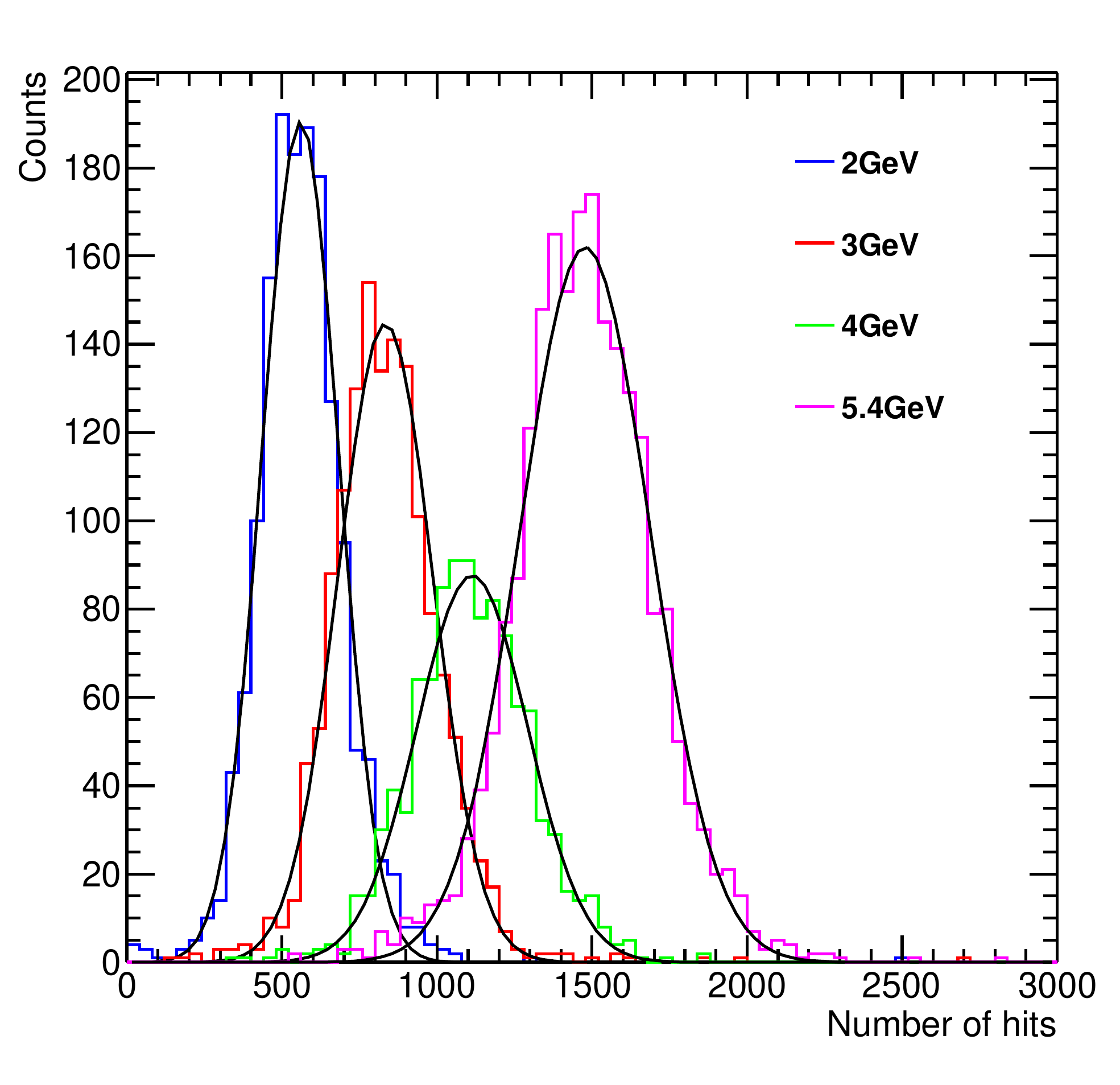}
\includegraphics[width=0.5\textwidth]{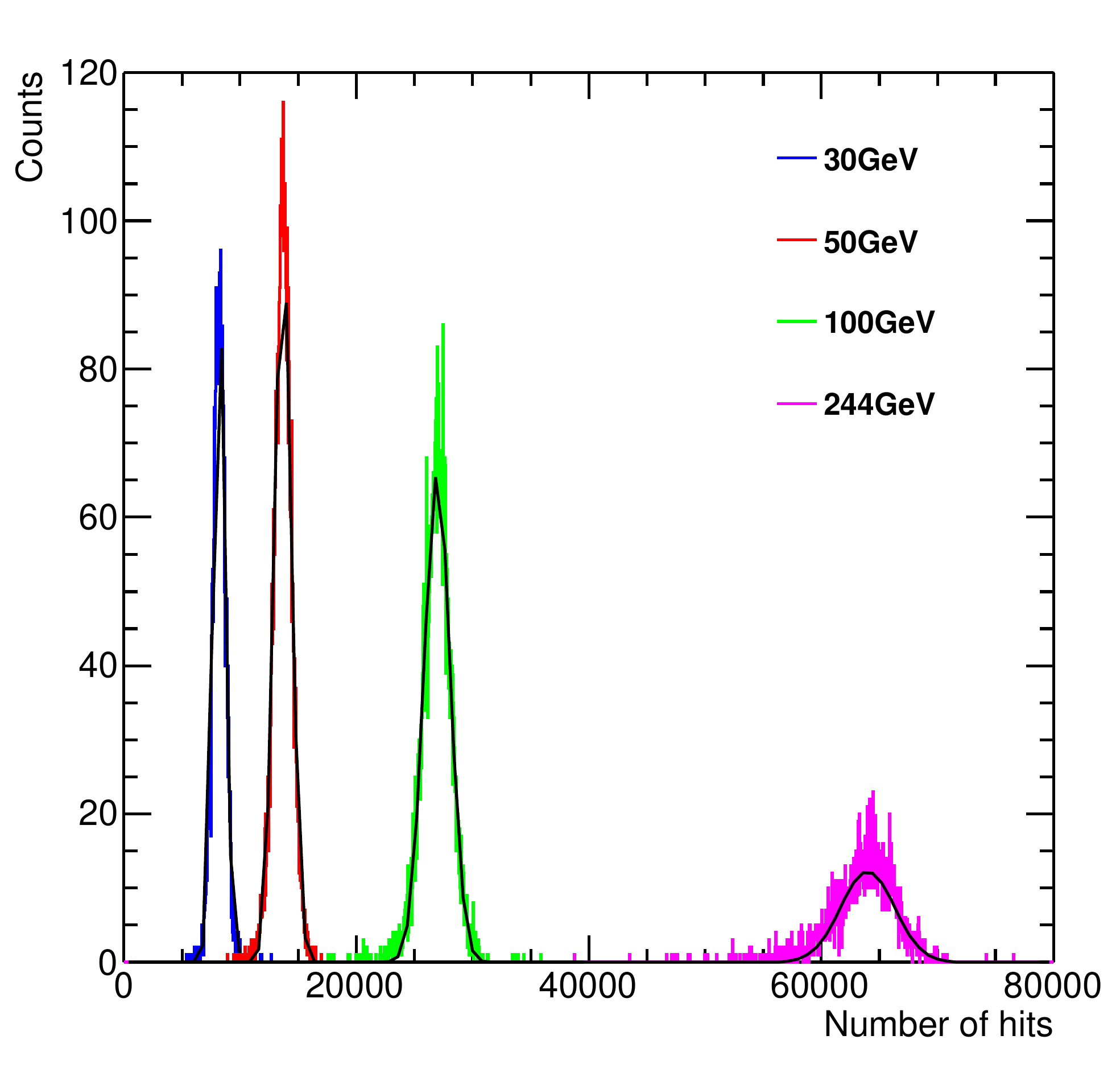}
\caption{Detector response (calibrated number of hits $K$) at low (left) and high (right) energies.}
\label{fig:response}
\end{figure}

The distributions of $K$ are shown for eight energies in figure~\ref{fig:response}.

\begin{figure}  [h]
\centering
\includegraphics[width=0.7\textwidth]{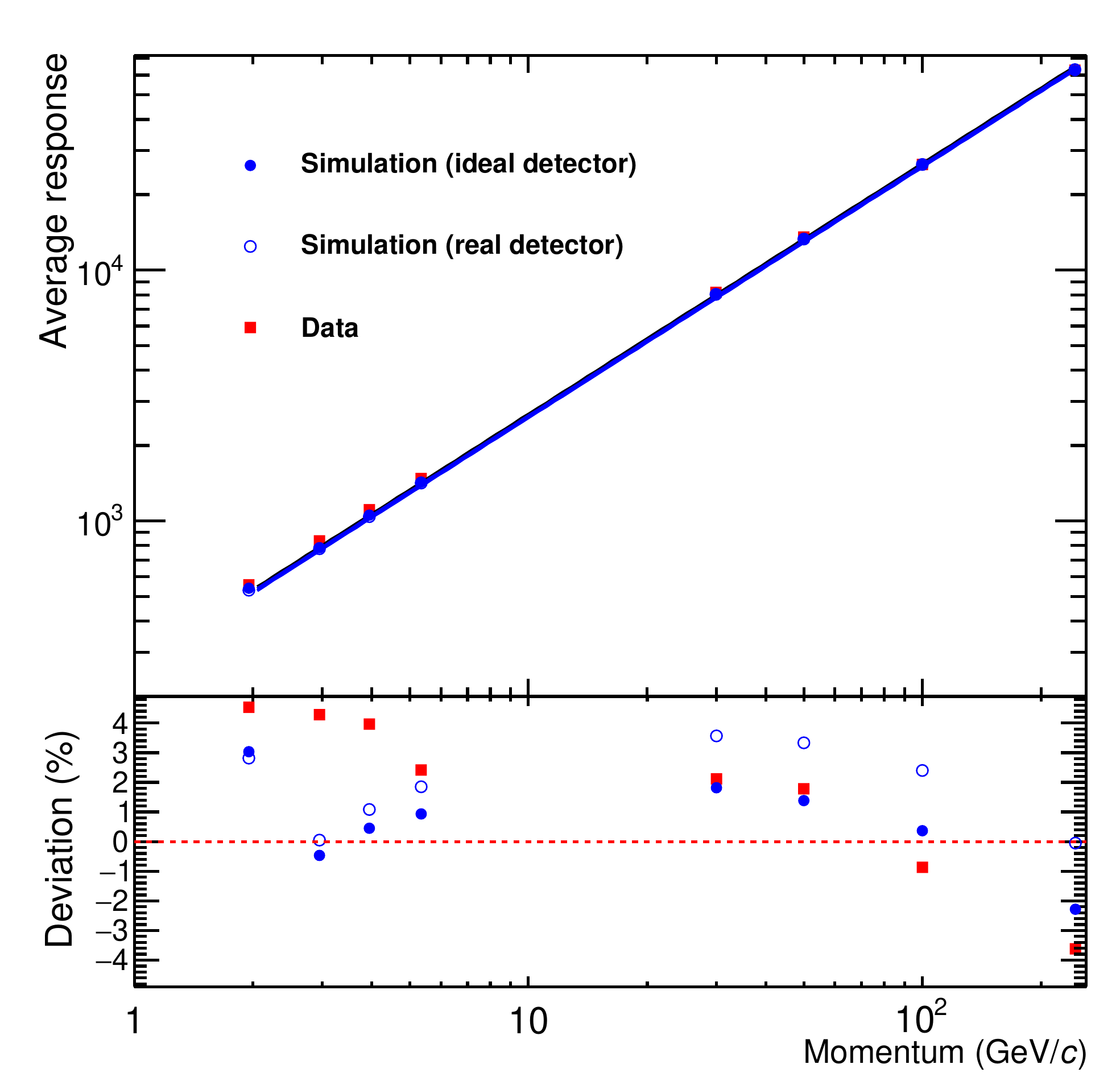}
\caption{Average detector response (top) and deviation from linearity (bottom) as a function of beam energy for electrons. In both panels the results from test beam data (red squares) are compared to simulations using two different implementations of the detector (see section~\ref{sec:simul}):  "ideal detector" (open blue circles) and "real" detector (full blue circles).}
\label{fig:linearity}
\end{figure}

The response (mean) and the resolution (width/mean) are obtained from a Gaussian fit.
The linearity of the response is shown in figure~\ref{fig:linearity}. The data at 2, 3, 4 and 5~GeV were taken with a slightly differently tuned detector, but calibrated in the same way as described in section~\ref{sec:calib}.

\begin{figure}  [h]
\centering
\includegraphics[width=0.7\textwidth]{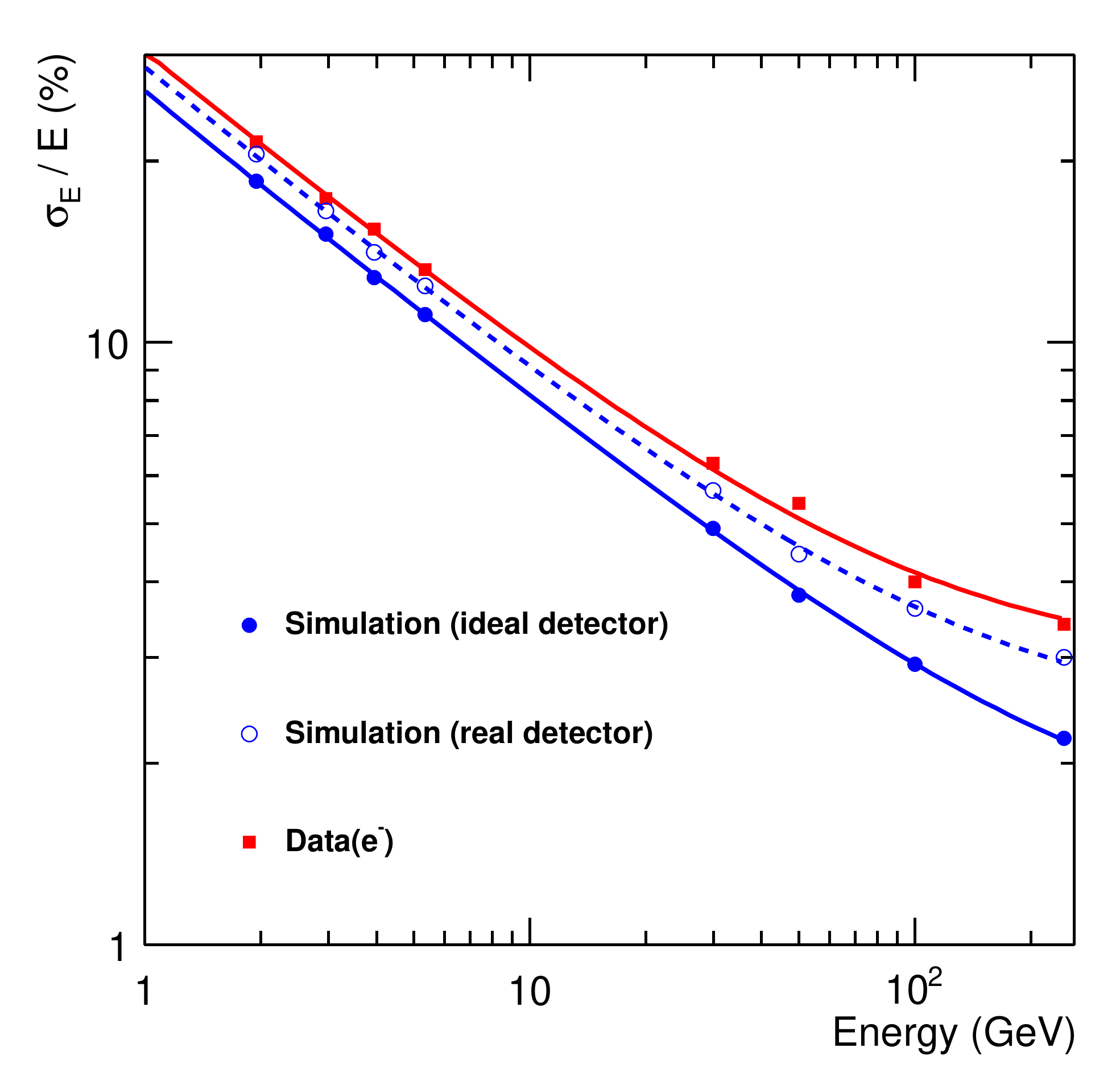}
\caption{Energy resolution as a function of beam energy for electrons. As in figure \protect\ref{fig:linearity} the data are compared to simulations using both the real and an ideal detector.}
\label{fig:resolution}
\end{figure}

The energy resolution is shown in figure~\ref{fig:resolution}.
The experimental data can be fitted with
\begin{equation}
 \frac{\sigma}{E} = A \oplus  \frac{B}{\surd E}  \oplus  \frac{C}{ E } \; ,
\end{equation}
with $A = (0.028 \pm 0.017)$, $B = (0.30 \pm 0.04)\sqrt{\mathrm{GeV}}$ and $C =  0.063 \; \mathrm{GeV}$.
The parameter $C$ is derived from the width of the pedestal peak (figure \ref{fig:noise}).

Both figures also show the results of two simulations. 
They are based on the procedure described in section~\ref{sec:simul}: an "ideal detector" which uses all sensors, a "real detector" which excludes all sensors and channels that are not working.
The first shows what would be ultimately achievable with this technique, given the size and segmentation of the prototype.
For the simulations the resulting parameters are 
$A = (0.015 \pm 0.018)$, $B = (0.25 \pm 0.03)\sqrt{\mathrm{GeV}}$ for the "ideal detector" and 
$A = (0.024 \pm 0.017)$, $B = (0.28 \pm 0.04)\sqrt{\mathrm{GeV}}$ for the "real detector".
The discrepancy between the second simulation and the experiment is partly due to the energy spread of the test beam (1.5\%).
Another cause may be the assumed homogeneity of the sensitivity of the sensors in the simulation.
As already shown in figure \ref{fig:mipcluster} left there is also a discrepancy in the simulation of the cluster size, which points in the same direction.

\begin{figure}  [tbh]
\centering
\includegraphics[width=0.8\textwidth]{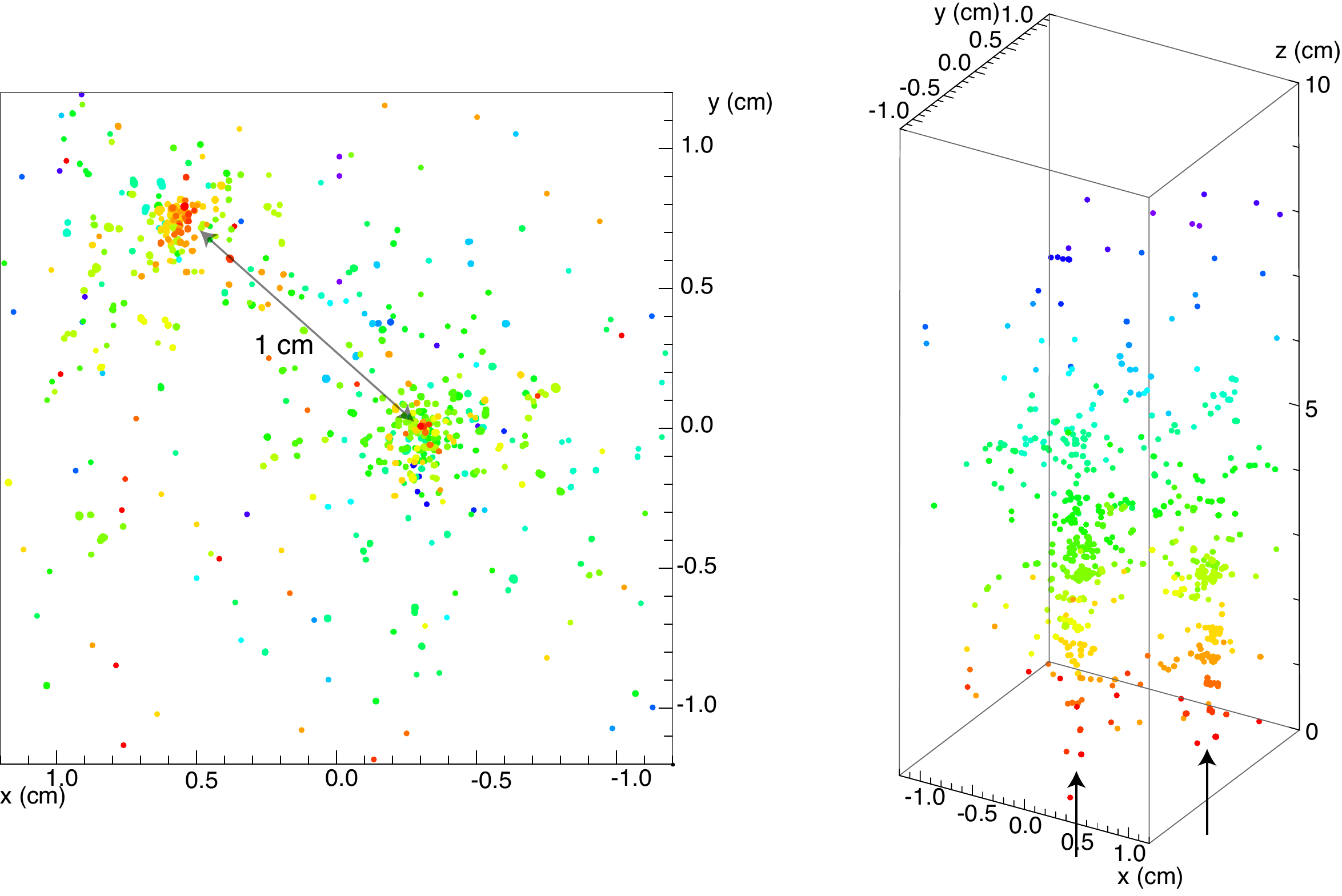}
\caption{Spatial distribution of hit pixels projected onto the transverse plane (left) and as seen from a side perspective (right). Shown is the simultaneous measurement of the showers of two electrons of 5.4~GeV/$c$. Every dot represents a single hit pixel. For illustration purposes the colour of the dots indicates the $z$ position of the layer, with red corresponding to the smallest and blue to the largest $z$-values. The two arrows indicate the approximate position of incidence of the two particles.}
\label{fig:event}
\end{figure}

The very narrow lateral distributions shown in figure~\ref{fig:lateral} suggest the possibility of using only the hits within a certain radius $R$ from the shower centre.
This is explored by applying an upper limit to $R$ in equation~\ref{eq:layerresponse}.
In this way one can retrieve information from near-by showers, like in figure \ref{fig:event}.

\begin{figure}  [h]
\centering
\includegraphics[width=0.4\textwidth]{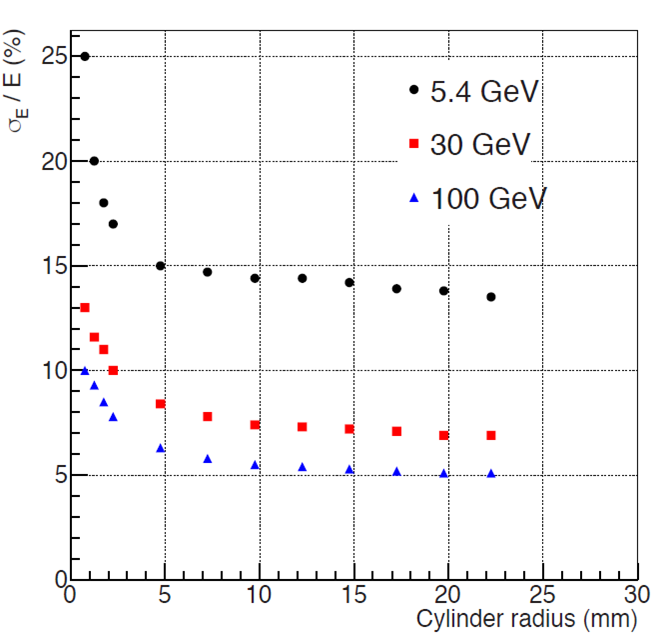}
\hspace{0.5 cm}
\includegraphics[width=0.4\textwidth]{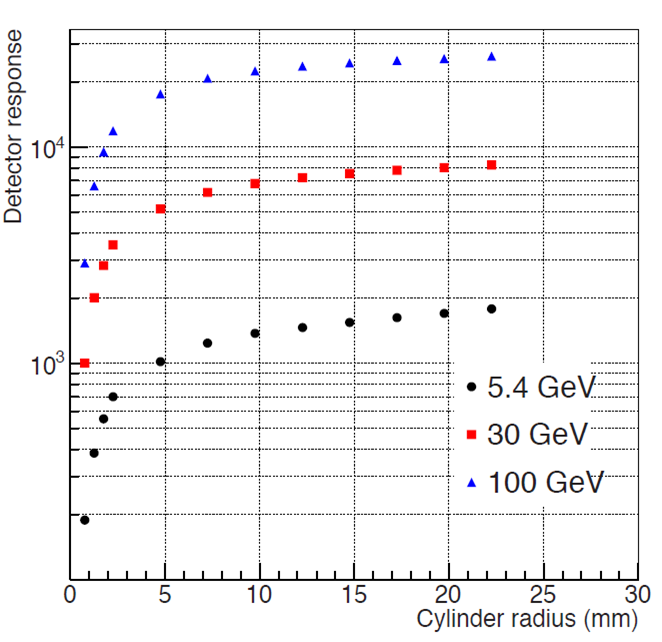}
\caption{Resolution and response as a function of the radius of the core for different energies.}
\label{fig:core}
\end{figure}

Figure~\ref{fig:core} shows the resolution and the response for a single shower as a function of this limiting radius $R$.
The resolution and the response are hardly affected down to half $R_\mathrm{M}$.

\section{Conclusion} \label{sec:concl}

A prototype EM calorimeter with fine sampling and pixel counting has been built and successfully tested with particle beams.
The prototype has a number of imperfections, most notably a large fraction of dead or otherwise unusable pixels, which can likely be improved in a new detector, but it nevertheless shows a very good performance.
Its small Moliere radius of 10~mm combined with the very fine sampling enables the direct separation of close showers,
using only the signal in a small core around the shower axis. This leads to a small deterioration of the resolution.
The position of a single electron shower can be determined with an accuracy of < 0.03~mm at high energy.
The data show the lateral profiles of showers in unprecedented detail.
The prototype shows an energy resolution of  
\begin{equation}
 \frac{\sigma}{E} = 2.8 \oplus  \frac{30}{\surd E\mathrm{(GeV)}}  \oplus  \frac{6.3}{ E\mathrm{(GeV)} } \%
\end{equation} 
which is reasonably well reproduced by simulation.
For a perfect detector, simulations show that a factor of two improvement can be expected.
Although at a few hundred GeV saturation in the shower core is observed, the non-linearity is only a few percent.
For energies at the TeV level, a smaller pixel size would be desirable.

The analysis of the prototype measurements clearly demonstrates the feasibility of such a highly granular digital calorimeter.
 The measurements constitute the basis for many more interesting studies, e.g. of:
\begin{itemize}
\item detailed properties of the three-dimensional distributions of electromagnetic showers and their event-by-event fluctuations, and
\item features of hadronic showers, including potentially: tracking within the shower, the distribution of the shower start, and electromagnetic components, i.e. information, which will be crucial for hadron rejection and the development of particle flow algorithms.
\end{itemize} 
The analysis will also allow to test and possibly improve the description of showers in MC simulation programs such as GEANT.
Such studies are beyond the scope of this paper and will likely be the subject of future publications.

\acknowledgments
We gratefully acknowledge the DESY and CERN accelerator staff for the reliable and efficient beam operations 
and the ALICE TC team for the general support for test beams at CERN. 
The authors would like to thank the Picsel group at IPHC Strasbourg for their assistance in acquiring the MIMOSA sensors.
The beam test at DESY in 2014 has received support of the European Community - Research Infrastructure Action under the FP7 \textquoteleft Capacities\textquoteright Specific Programme, AIDA-DESY-2014-04. 
This work was supported in part by the Chinese Scholarship Council and the Dutch research organisation NWO-I (formerly FOM).

%\clearpage

\end{document}